\documentclass[11pt,a4paper]{article}
\usepackage{jheppub}
\usepackage{textcomp}
\usepackage{amsmath, amssymb}
\usepackage{bbold}
\usepackage{mathbbol}
\usepackage{slashed}
\usepackage{graphics}
\usepackage{shadow}
\usepackage{float}
\usepackage{latexsym,feynmf}

\newcommand{\de}{\partial}
\newcommand{\braket}[2]{ \langle #1 \lvert #2 \rangle }  %braket
\newcommand{\ket}[1]{\lvert #1 \rangle} %ket
\newcommand{\bra}[1]{\langle #1 \rvert} %bra
\newcommand{\Braket}[3]{\langle #1 \lvert #3 \rvert #2 \rangle }  %complete braket
\newcommand{\pu}{\!\!\cdot}

\def\3i{\int\!\!\!\int\!\!\!\int}
\def\2i{\int\!\!\!\int}

\def\del{\Delta}
\def\ddel{{}^\bullet\! \Delta}
\def\deld{\Delta^{\hskip -.5mm \bullet}}
\def\dddel{{}^{\bullet \bullet} \! \Delta}
\def\ddeld{{}^{\bullet}\! \Delta^{\hskip -.5mm \bullet}}
\def\deldd{\Delta^{\hskip -.5mm \bullet \bullet}}
\def\ddelF{{}^\bullet\! \Delta_F}
\def\deldF{\Delta^{\hskip -.5mm \bullet}_F}
\title{Extended SUSY quantum mechanics: transition amplitudes and path integrals}
\author[a]{Fiorenzo Bastianelli,}
\author[a]{Roberto Bonezzi,}
\author[a,b]{Olindo Corradini,}
\author[c,d]{Emanuele Latini}

\affiliation[a] {Dipartimento  di Fisica, Universit{\`a} di Bologna and\\
INFN, Sezione di Bologna, via Irnerio 46, I-40126 Bologna, Italy}
\affiliation[b] {Centro de Estudios en F\'isica y Matem\'aticas Basicas y Aplicadas\\
Universidad Aut\'onoma de Chiapas, Tuxtla Guti\'errez, Chiapas, Mexico}
\affiliation[c] {Department of Mathematics, University of California,
            Davis CA 95616, USA}
\affiliation[d] {INFN, Laboratori Nazionali di Frascati, CP 13,
I-00044 Frascati, Italy}

\emailAdd{bastianelli@bo.infn.it}\emailAdd{bonezzi@bo.infn.it}\emailAdd{corradini@bo.infn.it}
\emailAdd{latini@lnf.infn.it}

\abstract{Quantum mechanical models with extended supersymmetry find interesting applications
in worldline approaches to relativistic field theories. In this paper we consider one-dimensional
nonlinear sigma models with O($N$) extended supersymmetry on the worldline, which are used
in the study of higher spin fields on curved backgrounds. We calculate the transition amplitude
for euclidean times (i.e. the heat kernel) in a perturbative expansion, using both canonical methods
and path integrals. The latter are constructed using three different regularization schemes, and the
corresponding counterterms that ensure  scheme independence are explicitly identified.}

\begin{document}
\maketitle
\flushbottom

%%%%%%%%%%%%%%%%%%%%%%%%%%%%%%%%%%%%
\begin{fmffile}{feynTamp}{
\fmfset{zigzag_width}{1thick}
\fmfset{dot_len}{0.75mm}
\fmfcmd{%
  style_def majorana expr p =
    cdraw p;
    cfill (harrow (reverse p, .25));
    cfill (harrow (p, .25))%;
    %cdrawdot p;
  enddef;}
\fmfcmd{vardef endpoint_dot expr p =
  save oldpen; pen oldpen;
  oldpen := currentpen;
  pickup oldpen scaled 4;
  cdrawdot p;
  pickup oldpen;
enddef;}
\fmfcmd{style_def deld expr p =
  cdraw p;
  %endpoint_dot point infinity of p;
  endpoint_dot point 0.25 of reverse p;
 enddef;}
\fmfcmd{style_def ddeld expr p =
  cdraw p;
  endpoint_dot point 0.25 of reverse p;
  endpoint_dot point 0.25 of p;
 enddef;}
\fmfcmd{style_def ddel expr p =
  cdraw p;
  endpoint_dot point 0.25 of p;
 enddef;}

%%%%%%%%%%%%%%%%%%%%%%%%%%%%%%%%%%%%%%%%%%%%%%%%%%%%%%%%%%%
\section{Introduction}
\label{sec:intro}
%%%%%%%%%%%%%%%%%%%%%%%%%%%%%%%%%%%%%%%%%%%%%%%%%%%%%%%%%%%
Quantum mechanical models with extended supersymmetry find useful applications
in the worldline description of relativistic field theories. Indeed fields with spin  $S$
can be described in four dimensions by  quantizing particle actions
with $N=2S$  extended local supersymmetry on the worldline \cite{Gershun:1979fb,Howe:1988ft,Howe:1989vn}.
While complete in four dimensions, these models describe only conformal invariant fields
in other dimensions  \cite{Siegel:1988ru,Siegel:1988gd}.
They can be consistenly defined not only in Minkowski space, but also in
maximally symmetric spacetimes  \cite{Kuzenko:1995mg}
and, more generally, in conformally flat spacetimes \cite{Bastianelli:2008nm}.
Upon gauge fixing, the worldline actions of the spinning particles
moving on such spaces give rise to an interesting class of one-dimensional
nonlinear sigma models possessing extended rigid supersymmetries. The goal of this paper
is to analyze the corresponding quantum mechanics.

In particular, we are interested in computing the transition amplitude
for the O($N$) extended supersymmetric nonlinear sigma models
in a euclidean short time expansion since, in applications to higher spin field theory, the transition amplitude
can be used to study ultraviolet properties of
propagators  and one-loop effective actions \cite{DeWitt:1985bc}.
We will achieve this result using two different methods.
The first method employs canonical quantization, and starting from the operatorial definition
of the transition amplitude  we compute it perturbatively in the euclidean time $\beta$
by using the commutation properties of the various operators.
The final result yields a perturbative solution of the heat equation (the Schr\"odinger equation with imaginary
time) and identifies a benchmark for equivalent calculations in terms of path integrals.
This canonical approach has been already employed in \cite{Peeters:1993vu}
for the $N=0,1,2$ supersymmetric quantum mechanics, that can be defined
on any curved manifold (see \cite{Bastianelli:2006rx}  for a review of the method in the bosonic case).
We extend that computation to arbitrary $N$ for the O($N$) extended supersymmetric nonlinear sigma models.
For $N>2$ the extended supersymmetry may be broken
by a generic curved target space, though it  can be maintained on locally symmetric spaces.
Nevertheless, we use an arbitrary target space geometry since for the present purposes
we do not need to gauge supersymmetry, and can view the latter
just as an accidental symmetry emerging on particular backgrounds.

The second method we employ for computing the transition amplitude makes use of path integrals.
Our reason for considering this approach is that,
in typical first-quantized applications, canonical methods are used first to identify the precise
operators of interest and path integrals are used next to perform more extensive calculations
and manipulations. A classical example is the calculation of chiral anomalies and the
proof of index theorems  by worldline methods \cite{Alvarez-Gaume:1983at,Alvarez-Gaume:1983ig,Friedan:1983xr}.
 It is therefore important to be able to reproduce  the transition amplitude with path integrals.
 Path integrals for particles moving on curved spaces can be quite subtle, and their consistent definition
 needs the specification of a regularization scheme which carry finite counterterms to guarantee
 scheme independence, see \cite{Bastianelli:2006rx} for an extensive treatment and \cite{Bastianelli:2005rc}
 for a short discussion.  We will use three different schemes, for completeness and because
 each one carries its own advantages. The first scheme, time slicing (TS),  can be deduced directly
 by using operatorial methods \cite{DeBoer:1995hv,deBoer:1995cb}
and can be related to the lattice approximation of the propagation time.
A second scheme, mode regularization (MR), is related to a momentum cut-off (or, more properly,
energy cut-off in quantum mechanics) and allows the introduction of a regulated functional space to integrate over
\cite{Bastianelli:1991be,Bastianelli:1992ct,Bastianelli:1998jm,Bastianelli:1998jb,Bonezzi:2008gs}.
Finally, dimensional regularization (DR) is defined as a  purely perturbative regularization but has the advantage
of carrying only covariant counterterms
\cite{Kleinert:1999aq,Bastianelli:2000pt,Bastianelli:2000nm,Bastianelli:2002qw,Bastianelli:2005vk}.
For each of the three regularization schemes we find the corresponding counterterms
that ensure scheme independence, making them ready for extending to curved backgrounds 
the worldline approach to higher spin fields initiated in \cite{Bastianelli:2007pv}. 

Let us now describe the precise form of the (supersymmetric) quantum mechanics we are interested in.
We consider a particle moving in a curved space ${\cal M}$ of dimensions $D$ and metric $g_{\mu\nu}$.
It is described in phase space by bosonic coordinates and momenta $(x^\mu,p_\mu)$,
where $\mu=1,\dots,D$ is a curved spacetime index,
%and by $N$ fermionic Majorana variables $\psi^a_i$,
and by Majorana worldline fermions (i.e. real Grassmann variables) $\psi^a_i$,
 where $i=1,\dots,N$ is a flavour index and $a=1,\dots,D$ is a flat
spacetime index related to curved  indices by the vielbein $e_\mu^a$, defined by
$g_{\mu\nu} = \eta_{ab}e_\mu^a e_\nu^b$ with $\eta_{ab}$
the flat tangent space metric
(which can be taken either minkowskian or euclidean, according to the desired applications).
 Quantum mechanically the bosonic variables satisfy the usual commutation relations
$[x^\mu, p_\nu]=i\delta^\mu_\nu$, and the fermionic ones give rise to a
multi Clifford algebra $\{\psi^a_i,\psi^b_j\}=\eta^{ab}\delta_{ij}$.
The general hamiltonian operator we wish to consider involves a free covariant kinetic term $H_0$,
a four-Fermi interaction depending on the Riemann tensor and carrying  a coupling constant $\alpha$,
and a contribution from an arbitrary scalar potential $V$ which
depends only on the spacetime  coordinates $x^\mu$ (and which in most applications
is proportional to the curvature scalar $R$)
\begin{equation}
H=H_0+\alpha R_{abcd} \psi^a_i\psi^b_i \psi^c_j\psi^d_j +V
\label{eq:H}
\end{equation}
where
\begin{eqnarray}
H_0 &=&\frac12\Big(\pi^a-i\omega_b{}^{ba}\Big)\pi_a \nonumber\\
\pi_a &=& e_a^\mu \pi_\mu \;, \quad \pi_\mu = g^{1/4}
p_\mu g^{-1/4}
-\frac{i}{2}\omega_{\mu ab} \psi^a_i\psi^b_i
\end{eqnarray}
with $\omega_{\mu ab}$ the spin connection and
$\omega_{abc} = e_a^\mu \omega_{\mu bc}$.
Note the appearance of the spin connection, required by covariance,
and the powers of $g$ next to the momentum operator that ensure hermiticity of the
 hamiltonian. This hamiltonian is general enough to allow future applications to
the first-quantized theory of higher spin fields on curved backgrounds.

 The corresponding euclidean classical action in configuration space is given by
\begin{eqnarray} \label{action1.3}
  && S=  \int_0^\beta d\tau\, \Big [
  \frac12 g_{\mu\nu} \dot x^\mu \dot x^\nu
+\frac{1}{2} \psi_{a i} D_\tau \psi^a_i   +\alpha
  R_{abcd}\psi^a_i\psi^b_i\psi^c_j\psi^d_j +V \Big ]
\end{eqnarray}
where $D_\tau \psi^a_i = \partial_\tau \psi^a_i
+ \dot x^\mu \omega_{\mu}{}^a{}_b \psi^b_i $.
It describes the particle propagation for a euclidean time $\beta$
 and will be used in the path integral quantization. For notational simplicity
 we do not make explicit distinction between quantum operators and classical variables,
 as it will be clear from the context which one is used.

 Having defined the model, we proceed with the rest of the paper and start
 describing the computation of the transition amplitude.
 In section 2 we use operatorial methods. In  section 3
 we consider path integrals in various regularization schemes,
 namely TS, MR and DR, calculating the corresponding counterterms and finding
 complete agreement with the expression of the transition amplitude found in section 2.
 Finally, we present our conclusions and perspectives in section 4.

%%%%%%%%%%%%%%%%%%%%%%%%%%%%%%%%%%%%%%%%%%%%%%%%%%%%%%%%%%%
\section{Transition amplitude from operatorial methods}
\label{sec:t-amp-op}

The aim of this section is the explicit computation of the transition amplitude for a euclidean time $\beta$
determined by the quantum mechanical hamiltonian $H$ given in (\ref{eq:H}).
We compute the matrix elements of the evolution operator ${e^{-\beta H}}$
between position eigenstates for the bosonic variables and suitable
coherent states for the fermionic ones. We consider a  short time expansion, and using systematically
the fundamental (anti)-commutation relations of the basic operators $x, p$ and $\psi$   we obtain
the final perturbative answer. The calculation is tailored after similar ones performed in
 \cite{Peeters:1993vu} for the $N=0,1,2$ supersymmetric quantum mechanics,
explained carefully in \cite{Bastianelli:2006rx} for the bosonic case, and in \cite{Bastianelli:2010ir}
for a class of supersymmetric sigma models on K\"ahler spaces which arise from the study 
of certain higher spin equations on complex manifolds \cite{Bastianelli:2009vj}.
As typical in semiclassical approximations, the result can be cast as the product of three terms:
i) the exponential of the classical action evaluated on the (perturbative) classical solution,
ii) a standard leading prefactor depending on the propagation time as $\beta^{-\frac{D}{2}}$, usually arranged
in the so-called Van Vleck determinant,
iii) a power series in positive powers of the propagation time, which identify the heat-kernel coefficients \cite{DeWitt:1985bc}.
In our calculation we keep the approximation up to the first non trivial heat kernel coefficient, i.e. up to order
$\beta$ in the power series, keeping in mind that the bosonic displacement
$z^\mu\equiv y^\mu-x^\mu$ can be considered of order $\sqrt{\beta}$, as explained later on.

In this section we restrict the calculation to even $N\equiv 2S$, with $S$ an integer, which allows us
to introduce complex combinations of the fermionic operators and consider the corresponding coherent
states as a (over)-complete basis of the associated Hilbert space.
This is appropriate for applications to fields of integer spin.
The method can be easily extended to the case of odd $N$, appropriate for applications
to fields with half-integer spin: one way is to introduce another set of auxiliary Majorana  fermions
(``doubling trick'') to be able to consider complex (Dirac) fermions and their coherent states.
However, we refrain from doing that at this stage, as we employ it in the path integral approach of the
following section.

Thus we consider $N=2S$, and introduce $S$ Dirac worldline fermions $\Psi^a_I$ 
out of the $2S$ Majorana fermions $\psi^a_i$ %with $i=1,...,2S$
\begin{equation}
\Psi^a_I=\frac{1}{\sqrt2}\,(\psi^a_I+i\psi^a_{I+S})\;,\quad
\bar\Psi^{aI}=\frac{1}{\sqrt2}\, (\psi^a_I-i\psi^a_{I+S})\,,\quad I=1,\dots,S
\end{equation}
so that the new variables display the usual creation-annihilation algebra
$$
\{\Psi^a_I,\bar\Psi^{bJ}\}=\eta^{ab}\delta^J_I\;.
$$
With these complex fields at hand we can readily construct coherent
states such that $\Psi^a_I\ket{\eta}=\eta^a_I\ket{\eta}$ and
$\bra{\bar\lambda}\bar \Psi^{aI}=\bra{\bar\lambda}\bar\lambda^{aI}$, with usual
normalization $\braket{\bar\lambda}{\eta}=e^{\bar\lambda\cdot\eta}$ (other
properties of fermionic coherent states can be found in
appendix~\ref{app:FCS}). Denoting by
$\ket{y\,\eta}\equiv\ket{y}\otimes\ket{\eta}$ where $\ket{y}$ is the usual
position eigenstate, the euclidean heat kernel we want to compute is
\begin{equation}\label{transition amplitude definition}
\Braket{x\,\bar\lambda}{y\,\eta}{e^{-\beta H}}
\end{equation}
where the quantum hamiltonian \eqref{eq:H} and the covariant momentum, written in terms of the new complex
variables, read
\begin{equation}\label{quantum H}
\begin{split}
H &= H_0+4\alpha R_{abcd}\,\bar\Psi^a\pu \Psi^b\,\bar \Psi^c\pu \Psi^d+V\\
\pi_\mu &= g^{1/4}(p_\mu-i\omega_{\mu ab}\bar\Psi^a\pu \Psi^b)g^{-1/4}
\end{split}
\end{equation}
with $\bar\Psi^a\pu\Psi^b\equiv\bar\Psi^{aI}\Psi^b_I$.
In close analogy with the procedure employed in \cite{Peeters:1993vu}
we first focus on the mixed amplitude containing momentum eigenstates on the right hand side
\begin{equation}\label{mixed amplitude}
\Braket{x\,\bar\lambda}{p\,\eta}{e^{-\beta
    H}}=\sum_{k=0}^\infty\frac{(-)^k}{k!}\,\beta^k\Braket{x\,\bar\lambda}{p\,\eta}{H^k}\;.
\end{equation}
From the above expansion in $\beta$ one could naively think to retain only the linear term, $e^{-\beta H}=1-\beta H+\mathcal{O}(\beta^2)$, to get the answer up to the desired order. This is not the case for nonlinear sigma models, as well-known (see \cite{Peeters:1993vu}).
In fact it is necessary to take into account contributions for every order $k$, the only truncation being the number of (anti)-commutators to keep track of.
In order to evaluate \eqref{mixed amplitude}, we push all $p$'s and $\Psi$'s in each factor
$H^k$ to the right hand side and all $x$'s and $\bar \Psi$'s to the left hand side, taking
into account all (anti)-commutators, and substitute with the
corresponding eigenvalues. Since the hamiltonian is quadratic in momenta, the matrix element of $H^k$ is a polinomial of degree $2k$ in $p$, so that in general one finds structures of the form
\begin{equation}\label{B coefficients defined}
\Braket{x\,\bar\lambda}{p\,\eta}{H^k}=\sum_{l=0}^{2k}B^k_l(x,\eta,\bar\lambda)p^l\braket{x}{p}
\braket{\bar\lambda}{\eta}\;,
\end{equation}
where $p^l$ stands for the part of such polinomial with precisely $l$-th powers of $p$ eigenvalues,
and the coefficients $B^k_l$ are computed in appendix~\ref{app:B-coeff}.
For plane waves we use the normalization
$$
\braket{x}{p}=(2\pi)^{-D/2}g^{-1/4}(x)\,e^{ip\cdot x}
$$
so, inserting in \eqref{transition amplitude definition} a momentum
completeness relation, $\mathbb{1}=\int d^Dp\, |p\rangle\langle p| %\ketbra{p}{p}
$,
and rescaling momenta by $p_\mu=\frac{1}{\sqrt{\beta}}\,q_\mu$, we
obtain for the transition amplitude
\begin{equation}
\begin{split}
\Braket{x\,\bar\lambda}{y\,\eta}{e^{-\beta H}} &=
(4\pi^2\beta)^{-D/2}[g(x)g(y)]^{-1/4}\int d^D\!q\,
e^{\frac{i}{\sqrt{\beta}}q\cdot(x-y)}e^{\bar\lambda\cdot\eta}\\
&\times\sum_{k=0}^\infty\frac{(-)^k}{k!}\sum_{l=0}^{2k}\beta^{k-l/2}B^k_l(x,\eta,\bar\lambda)\,q^l\;.
\end{split}
\end{equation}
Let us look at this formula: it is well-known that the leading free
particle contribution to \eqref{transition amplitude definition}
contains the exponential $\exp\{-\frac{(x-y)^2}{2\beta}\}$, that gives
the effective order of magnitude $(x-y)\sim\sqrt\beta$. Therefore, we
see that $q\sim\beta^0$ and it is immediate to realize that, for
all $k$, only the terms $B^k_{2k}$, $B^k_{2k-1}$ and $B^k_{2k-2}$ will
contribute up to order $\beta$.
Taking for $B^k_l$ the expressions given in  appendix~\ref{app:B-coeff}, it is
immediate to sum the series in $k$
\begin{align} \label{eq2.7}
&\Braket{x\,\bar\lambda}{y\,\eta}{e^{-\beta H}} = (4\pi^2\beta)^{-D/2}[g(x)g(y)]^{-1/4}\int d^D\!q\,e^{-\frac12 g^{\mu\nu}q_\mu q_\nu+\frac{i}{\sqrt{\beta}}q\cdot(x-y)}e^{\bar\lambda\cdot\eta}\,\Big\{1+\sqrt{\beta}\nonumber\\
&\times\Big[\frac{i}{2}g^\mu q_\mu
-\frac{i}{4}g^{\mu\nu\lambda}\,q_\lambda\,q_\mu\,q_\nu+ig^{\mu\nu}\omega_\mu
q_\nu\Big]+\beta\Big[-\frac{1}{32}\de_\mu\ln g\de^\mu\ln
g-\frac18\de_\mu\de^\mu\ln g\nonumber\\&-\frac18g^\mu\de_\mu\ln
g -\Big(\frac14\de^\mu g^\nu+\frac18g^\mu g^\nu+\frac18g^\sigma
g^{\mu\nu}_\sigma+\frac18g^{\mu\nu\sigma}_\sigma\Big)\,q_\mu\,q_\nu+\Big(\frac{1}{12}
g^{\mu\nu\sigma\lambda}+\frac18g^{\sigma\lambda\mu}g^\nu\nonumber\\&+\frac{1}{12}g^{\kappa\sigma\lambda}
g^{\mu\nu}_\kappa
+\frac{1}{24}g^{\sigma\lambda}_\kappa
g^{\mu\nu\kappa}\Big)\,q_\sigma\,q_\lambda\,q_\mu\,q_\nu-\Big(\frac{1}{32}g^{\mu\nu\lambda}
g^{\sigma\rho\kappa}\Big)\,q_\lambda\,q_\mu\,q_\nu\,
q_\kappa\,q_\sigma\,q_\rho\nonumber\\& -\frac12\de^\nu(g^{\lambda\sigma}\omega_\lambda)q_\nu\,q_\sigma
-\frac14g^{\lambda\sigma\mu}\omega_\mu\, q_\lambda q_\sigma-\frac12g^{\mu\nu}\omega_\mu\, g^{\sigma}\,q_\nu q_\sigma+\frac14 g^{\mu\nu}\omega_\mu\,g^{\alpha\beta\sigma}\,q_\nu q_\sigma q_\alpha q_\beta\nonumber\\
&-\frac14 g^{\mu\nu}\omega_\mu \de_\nu\ln g+\frac{1}{2\sqrt
  g}\de_\mu[\sqrt g g^{\mu\nu}
  \omega_\nu]+\frac12(g^{\mu\nu}\omega_{\mu ab}\omega_{\nu
  cd}-8\alpha\, R_{abcd})(\bar\lambda^a\pu\eta^d\eta^{bc}\nonumber\\
&+\bar\lambda^a\pu\eta^b\bar\lambda^c\pu\eta^d)-\frac12g^{\mu\nu}\omega_{\mu
  ab}g^{\lambda\sigma}\omega_{\lambda
  cd}(\bar\lambda^a\pu\eta^d\eta^{bc}+\bar\lambda^a\pu\eta^b\bar\lambda^c\pu\eta^d)q_\nu
q_\sigma  - V \Big]\Big\}\;,
\end{align}
where we remind that all the geometric quantities are evaluated at the final point $x$, unless otherwise specified, and we used the following compact notations
\begin{equation}\nonumber
\omega_\mu=\omega_{\mu ab}\bar\lambda^a\pu\eta^b\;,\ \
g^{\nu\sigma}_{\mu...\lambda}=\de_\mu...\de_\lambda g^{\nu\sigma}\;,\ \
g^{\lambda\sigma\mu}= g^{\mu\nu}g^{\lambda\sigma}_\nu\;,\ \
g^\mu = g^{\mu\nu}_\nu\;, \ \
\de^\mu g^\lambda =g^{\mu\nu}\de_\nu g^{\lambda\sigma}_\sigma\;.
\end{equation}
Now we are ready to perform the $q$ integration, that reduces to a
bunch of gaussian integrals. Defining the coordinate displacement as
$z^\mu \equiv y^\mu-x^\mu$, one obtains the transition amplitude,
expanded up to order $\beta$, in a  form that hides manifest
covariance
\begin{align}
&\Braket{x\,\bar\lambda}{y\,\eta}{e^{-\beta H}}=
  (2\pi\beta)^{-D/2}\,\big[g(x)/g(y)\big]^{1/4}\,e^{-\frac{1}{2\beta}g_{\mu\nu}z^\mu z^\nu}\,
e^{\bar\lambda_a\cdot\eta^a}\,\Big\{1+z^\mu\,g^{-1/4}\,\de_\mu\,g^{1/4}\nonumber\\
&-\frac{1}{4\beta}\,\de_\lambda
g_{\mu\nu}\,z^\mu z^\nu z^\lambda+\frac12\,z^\mu z^\nu\,g^{-1/4}\,\de_\mu
\de_\nu g^{1/4}
-\frac{1}{4\beta}\,z^\mu g^{-1/4}\de_\mu g^{1/4}\de_\lambda
g_{\sigma\rho}\,z^\lambda z^\sigma z^\rho\nonumber\\
&+\frac12\,\Big[\frac{1}{4\beta}\,\de_\lambda
  g_{\mu\nu}\,z^\mu z^\nu z^\lambda\Big]^2-\frac{1}{12\beta}
\,\Big[\de_\lambda\de_\sigma
  g_{\mu\nu}-\frac12g_{\rho\tau}\,\Gamma^\rho_{\mu\nu}\,
  \Gamma^\tau_{\lambda\sigma}\Big]z^\mu z^\nu z^\lambda z^\sigma\nonumber\\&+\frac{1}{12}\,
R_{\mu\nu}\,z^\mu z^\nu
+z^\mu\omega_\mu+\frac12z^\mu z^\nu\de_\mu\omega_\nu+\frac14z^\mu\omega_\mu\,
z^\nu g^{\lambda\sigma}\de_\nu
g_{\lambda\sigma}\nonumber\\&+\frac12z^\mu z^\nu\omega_{\mu
  ab}\omega_\nu{}^b{}_c\bar\lambda^a\pu\eta^c+\frac12(z^\mu\omega_\mu)^2-z^\mu\omega_\mu\,
\Big(\frac{1}{4\beta}z^\nu z^\lambda z^\sigma\de_\nu
g_{\lambda\sigma}\Big)\nonumber\\ &+\beta\Big[-4\alpha\,R_{abcd}\bar\lambda^a\pu\eta^b\bar\lambda^c\pu\eta^d
  +4\alpha
  \,R_{ab}\bar\lambda^a\pu\eta^b+\frac{1}{12}R-V\Big]\Big\}  \;.
\label{tae}
\end{align}
This form is quite explicit. However, it can be simplified and cast in alternative and more suggestive forms.
Keeping in mind that the exponent of the on-shell action should appear in the final result,
one can factorize \eqref{tae} in the following way\footnote{The factor
$[g(x)/g(y)]^{1/4}$ cancel against its inverse, whose
Taylor expansion around $x$ can be factored out from \eqref{tae}.}
\begin{align}
&\Braket{x\,\bar\lambda}{y\,\eta}{e^{-\beta
H}}=(2\pi\beta)^{-D/2}
\nonumber\\ &
\times
\exp\Big\{\!\!-\frac{1}{\beta}\,\Big[\frac12\,g_{\mu\nu}\,z^\mu z^\nu+\frac14\,\de_\mu g_{\nu\lambda}\,z^\mu z^\nu z^\lambda
+\frac{1}{12}\,
\Big(\de_\mu\de_\nu g_{\lambda\sigma}
-\frac12\,g_{\rho\tau}\,\Gamma^\rho_{\mu\nu}\,\Gamma^\tau_{\lambda\sigma}\Big)\,z^\mu z^\nu z^\lambda z^\sigma\Big]\Big\}\nonumber\\
&\times\exp\Big\{\bar\lambda^a\pu\eta_a+z^\mu\omega_{\mu ab}\bar\lambda^a\pu\eta^b+\frac12z^\mu z^\nu\Big(\de_\mu\omega_{\nu ab}+\omega_{\mu ac}\omega_\nu{}^c{}_b\Big)\bar\lambda^a\pu\eta^b
-4\alpha\beta\, R_{abcd}\bar\lambda^a\pu\eta^b\bar\lambda^c\pu\eta^d\Big\}
\nonumber\\&\times\Big[1+\frac{1}{12}R_{\mu\nu}z^\mu z^\nu+\beta\Big(4\alpha R_{ab}\bar\lambda^a\pu\eta^b+\frac{1}{12}R - V\Big)\Big] \;.
\label{transition amplitude factorized}
\end{align}
The first exponential is the expansion up to order $\beta$ of the on-shell bosonic action
evaluated with the boundary conditions $x^\mu(0)=y^\mu$ and $x^\mu(\beta)=x^\mu$
\begin{equation}
\begin{split}
S_x &= \int_0^\beta d\tau\,\Big[\frac12 g_{\mu\nu}\dot x^\mu\dot x^\nu\Big]_{\text{on shell}}\\
&= \frac{1}{\beta}\,\Big[\frac12\,g_{\mu\nu}\,z^\mu z^\nu+\frac14\,\de_\mu g_{\nu\lambda}\,z^\mu z^\nu z^\lambda+\frac{1}{12}\,\Big(\de_\mu\de_\nu g_{\lambda\sigma}-\frac12\,g_{\rho\tau}\,\Gamma^\rho_{\mu\nu}\,\Gamma^\tau_{\lambda\sigma}\Big)\,z^\mu z^\nu z^\lambda z^\sigma\Big]
\end{split}
\end{equation}
where all functions in the second line are evaluated at the point $x$.
Similarly, one can see that the second exponential
in \eqref{transition amplitude factorized} is the expansion, up to
order $\beta$, of the fermionic action evaluated
on-shell, with boundary conditions $\Psi^a_I(0)=\eta^a_I$ and
$\bar \Psi^{aI}(\beta)=\bar\lambda^{aI}$, and with the usual boundary term
added
\begin{equation}
\begin{split}
 S_\Psi &= \left(\int_0^\beta d\tau\,\Big[\bar
  \Psi_a^ID_\tau\Psi^a_I +4\alpha\,R_{abcd}\bar \Psi^a\pu
  \Psi^b\bar \Psi^c\pu \Psi^d\Big]-\bar \Psi_a(\beta)\cdot \Psi^a(\beta)\right)\Bigg\rvert_{\text{on shell}}\\
&=-\Big\{\bar\lambda^a\pu\eta_a+z^\mu\omega_{\mu ab}\bar\lambda^a\pu\eta^b+\frac12z^\mu z^\nu\Big(\de_\mu\omega_{\nu ab}+\omega_{\mu ac}\omega_\nu{}^c{}_b\Big)\bar\lambda^a\pu\eta^b
-4\alpha\beta\, R_{abcd}\bar\lambda^a\pu\eta^b\bar\lambda^c\pu\eta^d\Big\}
\end{split}
\end{equation}
where the covariant time derivative reads
$D_\tau\Psi^a_I=\dot \Psi^a_I+\dot x^\mu\omega_\mu{}^a{}_b\,\Psi^b_I$.
Similar calculations of on-shell actions can be found, for instance, in  \cite{Peeters:1993vu}.
Once the expansions of the on-shell classical actions have been recognized, the transition
amplitude can be cast in the following covariant form
\begin{equation} 
\begin{split}
\Braket{x\,\bar\lambda}{y\,\eta}{e^{-\beta H}}
&= \frac{1}{(2\pi\beta)^{D/2}}\,\exp\left\{-(S_x+ S_\Psi)\right\}\\
&\times\Big[1+\frac{1}{12}R_{\mu\nu}z^\mu z^\nu+\beta\Big(4\alpha R_{ab}\bar\lambda^a\pu\eta^b
+\frac{1}{12}R - V \Big)+\mathcal{O}(\beta^2)\Big]
\end{split} \label{final result TA}
\end{equation}
with all functions evaluated at  point $x$.
This is our final result for the transition amplitude.

For comparison purposes, it may be useful to present the result after tracing over the fermionic  Hilbert space
\begin{equation} \begin{split}
\Braket{x}{y}{{\rm Tr}_\Psi (e^{-\beta H})} &= \frac{2^\frac{N D}{2}}{(2\pi\beta)^{D/2}}\,\exp\left\{-S_x \right\}
\Big[ 1- \frac{N}{16} z^\mu z^\nu  \omega_{\mu ab} \omega_{\nu}{}^{ab} \Big]\\
&\times  \Big [1+\frac{1}{12}R_{\mu\nu}z^\mu z^\nu+ \beta
\Big( \frac{1+6\alpha N}{12} R - V\Big )
+\mathcal{O}(\beta^2)\Big] \;.
\end{split}
\end{equation}
Evaluated at coinciding points ($z^\mu=0$) it reads as
\begin{equation} \begin{split} \label{traced TA-coinciding points}
\Braket{x}{x}{{\rm Tr}_\Psi (e^{-\beta H})} &= \frac{2^\frac{N D}{2}}{(2\pi\beta)^{D/2}}\,
\Big[1+\beta\Big( \frac{1+6\alpha N}{12} R - V\Big )+\mathcal{O}(\beta^2)\Big]
\end{split}
\end{equation}
which shows the first heat kernel coefficient at coinciding points for our model.
As we shall see, the last two formulas remain valid also for odd $N$.

%%%%%%%%%%%%%%%%%%%%%%%%%%%%%%%%%%%%%%%%%%%%%%%%%%%%%%%%%%%
\section{Transition amplitude from path integrals}
\label{sec:t-amp}
%%%%%%%%%%%%%%%%%%%%%%%%%%%%%%%%%%%%%%%%%%%%%%%%%%%%%%%%%%%
In the present section we compute the transition amplitude by making use of path integral methods.
To define the path integrals we fix a suitable regularization scheme, and
compute the transition amplitude. Then, by comparing with the previous section or
alternatively by imposing the Schr\"odinger equation, we identify the corresponding counterterms
 that ensure scheme independence.

Unlike the previous section here we treat together both the cases with even
and odd numbers of Majorana variables.
In order to do so, we found it convenient to use the so-called ``doubling trick'' that consists in
doubling the number of fermionic variables by adding ``spectator'' Majorana
fermions $\psi'^a_i$ that satisfy free anticommutation relations.
These new fermions are spectators in that they do not enter the interactions.
With the help of these new variables one can consider Dirac fermions
\begin{equation}
  \Psi^a_i= \frac1{\sqrt2}\Big(\psi^a_i+i\psi'^a_i\Big)\,,\quad
  \bar\Psi^{a}_i= \frac1{\sqrt2}\Big(\psi^a_i-i\psi'^a_i\Big)
\end{equation}
that satisfy
\begin{equation}
\{\Psi^a_i,\bar\Psi^{b}_j\} = \eta^{ab} \delta_{ij}
\end{equation}
giving rise to a set of fermionic harmonic oscillators, whose Hilbert
space can be represented in terms of bra and ket coherent states
\begin{eqnarray}
  \Psi^a_i|\eta\rangle = \eta^a_i|\eta\rangle\,,\quad
  \langle \bar\lambda| \bar\Psi^{a}_i = \langle \bar\lambda| \bar\lambda^{a}_i
\end{eqnarray}
and whose properties are briefly recalled in appendix \ref{app:FCS}. The wave function of
the system $\Phi(x,\bar\lambda) =  \langle x\,\bar\lambda| \Phi \rangle$, with
$\langle x\,\bar\lambda| \equiv \langle x|\otimes \langle \bar\lambda| $,
evolves in euclidean time as
\begin{eqnarray}
 \Phi(x,\bar\lambda;\beta)
 &=&
 \langle x\,\bar\lambda|e^{-\beta H}|\Phi\rangle
 \nonumber\\
 % &=&\int d^Dy \sqrt{g(y)}\int d\bar\zeta d\eta\ e^{-\bar\zeta\cdot \eta}
  %\langle x\,\bar\lambda|e^{-\beta H}|y\,\eta\rangle\langle y\,\bar\zeta|\Phi \rangle\nonumber\\
  &=&\int d^Dy \sqrt{g(y)}\int d\bar\zeta d\eta\ e^{-\bar\zeta\cdot \eta}
  \langle x\,\bar\lambda|e^{-\beta H}|y\,\eta\rangle\Phi(y,\bar\zeta;0)
\label{eq:evolution}
\end{eqnarray}
where we have used the spectral decomposition of unity for the
bosonic and fermionic sectors
\begin{equation}
  {\mathbb 1}_b= \int d^Dy \sqrt{g(y)} |y\rangle\langle y|\ ,\quad
  {\mathbb 1}_f= \int d\bar\zeta d\eta\ e^{-\bar\zeta\cdot \eta}
  |\eta\rangle\langle \bar\zeta|~.
\end{equation}
The evolution is generated by the basic
transition amplitude that can be written in terms of a path
integral as
\begin{eqnarray}
  \langle x\,\bar\lambda|e^{-\beta H}|y\,\eta\rangle
  &=&
  \langle x\,\bar\lambda;\beta|y\,\eta;0\rangle
  \nonumber\\
  &=&\int_{x(-1)=y}^{x(0)=x}DxDaDbDc
  \int_{\Psi(-1)=\sqrt{\beta}\eta}^{\bar\Psi(0)=\sqrt{\beta}\bar\lambda}D\bar\Psi
  D\Psi\ e^{-S[x,a,b,c,\Psi,\bar\Psi]}
\end{eqnarray}
where
\begin{eqnarray}
  && S[x,a,b,c,\Psi,\bar\Psi]=\frac1{2\beta}\int_{-1}^0d\tau\,
  g_{\mu\nu}(x(\tau)) (\dot x^\mu \dot x^\nu +a^\mu a^\nu +b^\mu
  c^\nu)\nonumber\\ &&+
  \frac1{\beta} \int_{-1}^0d\tau\, \bar\Psi_{ai}
  \dot\Psi^a_i -\frac1{\beta} \bar\Psi_{ai}(0)
    \Psi^a_i(0)+\frac1{2\beta}\int_{-1}^0d\tau\, \dot x^\mu
  \omega_{\mu ab}(x(\tau))\psi^a_i\psi^b_i\nonumber\\
  &&+\frac{\alpha}{\beta}\int_{-1}^0d\tau\,
  R_{abcd}(x(\tau))\psi^a_i\psi^b_i\psi^c_j\psi^d_j +\beta   \int_{-1}^0d\tau\, V(x(\tau))~.
\label{eq:S}
\end{eqnarray}
Here, and in the following, we use a shifted and rescaled euclidean time $\tau \in [-1,0]$ to make
comparison with the literature easier. We use bosonic ($a^\mu$)
and fermionic ($b^\mu,\, c^\mu$) ghosts to reproduce
the reparametrization invariant measure ${\cal D}x =\prod_\tau \sqrt{g(x(\tau))} d^Dx(\tau)$.
We also rescaled fermionic ``coordinates'' and ghosts so that all propagators are of order
$\beta$, and added a fermionic boundary term to be able to set
boundary conditions
at initial time for $\Psi$ and at final time for $\bar\Psi$.
Finally, let us note
that the arbitrary potential $V$  will eventually be modified by the addition of a counterterm  $V_{CT}$
related  to the regularization chosen. Apart form these modifications, 
and the inclusion of the spectator fermions,
this is the same action given in eq. (\ref{action1.3}).

We are interested in the short-time perturbative expansion of the
transition amplitude. Thus we expand all geometric expressions about
the final point $x^\mu$, and split the action into a quadratic part plus
interactions, $S=S_2+S_{int}$, with
\begin{eqnarray}
  S_2= \frac1{2\beta}g_{\mu\nu}\int_{-1}^0d\tau\, (\dot x^\mu \dot x^\nu +a^\mu a^\nu +b^\mu c^\nu)+
  \frac1{\beta} \int_{-1}^0d\tau\, \bar\Psi_{ai}
  \dot\Psi^a_i -\frac1{\beta} \bar\Psi_{ai} (0)\Psi^a_i(0)~.
  \label{eq:S2}
\end{eqnarray}
Here and in the following, geometric quantities with no explicit functional
dependence are meant to be evaluated at the final point $x(0)=x$, such as $g_{\mu\nu}
= g_{\mu\nu}(x)$. We can thus split the fields into backgrounds,
satisfying the free equations of motion with corresponding boundary conditions, and quantum fluctuations;
namely
\begin{align}
 & x^\mu(\tau) = \tilde x^\mu(\tau) + q^\mu(\tau)\,,
 &
 &\tilde  x^\mu(\tau) = x^\mu- z^\mu\tau\,,
 &
 & q^\mu(0)=q^\mu(-1)=0    \label{bc-uno}\\
  &\Psi^a_i(\tau) =\tilde\Psi^a_i(\tau) + Q^a_i(\tau)\,,
  &
  &\tilde\Psi^a_i(\tau)=\sqrt{\beta}\eta^a_i \,
  &
  &Q^a_i(-1) =0\label{eq:P}\\
  &\bar\Psi^a_i(\tau) =\tilde{\bar\Psi}^a_i(\tau) + \bar Q^a_i(\tau)\,,
  &
    &\tilde{\bar\Psi}^a_i(\tau)=\sqrt{\beta}\bar\lambda^a_i\,,
    &
    &\bar Q^a_i(0)
   =0 \label{eq:barP}
\end{align}
where $ z^\mu\equiv y^\mu-x^\nu$.
The free on-shell classical action reads
(henceforth we use a dot to indicate the contraction of whatever type of free flat indices)
\begin{equation}
  \tilde S_2 = \frac1{2\beta} g_{\mu\nu} z^\mu z^\nu-\bar\lambda\cdot \eta
\end{equation}
and the free propagators for the quantum fields,
together with their Feynman diagrams, are
\begin{eqnarray}
  \langle q^\mu(\tau)q^\nu(\sigma)\rangle &=&-\beta
  g^{\mu\nu}\Delta(\tau,\sigma)=\quad
      \parbox{50pt}{
        \begin{fmfgraph}(50,15)
         %\fmfpen{thick}
         \fmfleft{i} \fmfright{o} \fmf{plain}{i,o}
      \end{fmfgraph}}
      \label{qq}
  \\
  \langle a^\mu(\tau) a^\nu(\sigma)\rangle &=& \beta
  g^{\mu\nu}\Delta_{gh}(\tau,\sigma)=\quad
      \parbox{50pt}{
        \begin{fmfgraph}(50,15)
         %\fmfpen{thick}
         \fmfleft{i} \fmfright{o} \fmf{dashes}{i,o}
      \end{fmfgraph}}
  \label{aa}
  \\
  \langle b^\mu(\tau) c^\nu(\sigma)\rangle &=& -2\beta
  g^{\mu\nu}\Delta_{gh}(\tau,\sigma)=\quad
      \parbox{50pt}{
        \begin{fmfgraph}(50,15)
          %\fmfpen{thick}
          \fmfleft{i} \fmfright{o} \fmf{dashes}{i,o}
      \end{fmfgraph}}
  \label{bc}
  \\
  \langle \bar Q^a_i(\tau) Q^b_j(\sigma)\rangle &=&\beta
  \eta^{ab}\delta_{ij} G(\tau,\sigma)
  \label{eq:QQ}
\end{eqnarray}
where the right-hand sides are given, at the unregulated level, by the following distributions
\begin{equation}\label{eq:D}
\begin{split}
\Delta(\tau,\sigma) &= \tau(\sigma+1)\theta(\tau-\sigma)+\sigma(\tau +1)\theta(\sigma-\tau)\\
\Delta_{gh}(\tau,\sigma)&= \delta(\tau-\sigma)\\
G(\tau,\sigma)&= -\theta(\sigma-\tau)
\end{split}
\end{equation}
which obey the Green equations
$\dddel(\tau,\sigma) =\Delta_{gh}(\tau,\sigma)=\delta(\tau-\sigma)$
and  ${}^{\bullet}G(\tau,\sigma) =\delta(\tau-\sigma)$,
with  boundary conditions
$\Delta(0,\sigma)=\Delta(\tau,0)=\Delta(-1,\sigma) =\Delta(\tau,-1)=0$
and $G(0,\sigma) = G(\tau,-1)=0$.
Dots to the left (right) indicate derivatives with respect to the first (second) variable.
These propagators have to be regulated in order to
deal with divergences and ambiguities present in some diagrams. However,
for each regularization scheme, one is
able to cast all expressions in a way that can be unambiguously
computed by using directly the expressions \eqref{eq:D}.

Since fermions enter the interactions only through the
combination $\psi^a_i=\frac1{\sqrt2}(\Psi^a_i+\bar\Psi^a_i)$ it is
convenient to write backgrounds and propagators for these fields as well. We
split them as $\psi^a_i(\tau) =\tilde\psi^a_i +\chi^a_i(\tau)$ with
\begin{eqnarray} \label{tilde fermion}
  \tilde\psi^a_i=\sqrt{\frac{\beta}2}(\eta^a_i+\bar\lambda^a_i)
\end{eqnarray}
and
\begin{eqnarray}
  \langle \chi^a_i(\tau)\chi^b_j(\sigma)\rangle
  &=&\beta\eta^{ab}\delta_{ij}\Delta_F(\tau,\sigma)=\quad
      \parbox{50pt}{
        \begin{fmfgraph}(50,15)
         %\fmfpen{thick}
         \fmfleft{i} \fmfright{o} \fmf{dots}{i,o}
\end{fmfgraph}}
\label{chichi}
\end{eqnarray}
satisfying
$\ddelF(\tau,\sigma)=-\deldF(\tau,\sigma)=\delta(\tau-\sigma)$
and given  at the unregulated level by
\begin{eqnarray}\label{eq:Df}
  \Delta_F(\tau,\sigma)=\frac12( \theta(\tau-\sigma)  -\theta(\sigma-\tau))
  =\frac12\epsilon(\tau-\sigma)\;.
\end{eqnarray}
We only wrote propagators for unprimed fermionic fields since
only they enter the interactions. The primed fields instead only contribute to
the one-loop semiclassical factor that normalizes the path
integral and drop immediately out of the computation.

The interaction vertices that enter the perturbative expansion can
be obtained by Taylor expanding the action~(\ref{eq:S}) about the
final point $x(0)=x$ and read $S_{int} = \sum_{n=3}^\infty S_n$, with
\begin{eqnarray}
S_3 &&=  \frac1{2\beta}\omega_{\mu ab} \int_{-1}^0d\tau
(\dot q^\mu - z^\mu)(\tilde\psi^a+\chi^a)\cdot(\tilde\psi^b+\chi^b)\\
&&+\frac1{2\beta}\partial_\lambda g_{\mu\nu} \int_{-1}^0d\tau
(q^\lambda- z^\lambda\tau) \Big(
(\dot q^\mu- z^\mu)(\dot q^\nu- z^\nu)
+a^\mu a^\nu +b^\mu c^\nu\Big)\nonumber\\
S_4 &&= \beta V + \frac1{2\beta}\partial_\lambda\omega_{\mu ab}
\int_{-1}^0d\tau(q^\lambda- z^\lambda\tau)(\dot q^\mu- z^\mu)
(\tilde\psi^a+\chi^a)\cdot(\tilde\psi^b+\chi^b)\\
&&+\frac{\alpha}{\beta}R_{abcd}\int_{-1}^0d\tau
(\tilde\psi^a+\chi^a)\cdot(\tilde\psi^b+\chi^b)(\tilde\psi^c+\chi^c)\cdot(\tilde\psi^d+\chi^d)
\nonumber\\
&&+ \frac1{4\beta}\partial_\lambda\partial_\sigma g_{\mu\nu} \int_{-1}^0d\tau
(q^\lambda- z^\lambda\tau)(q^\sigma- z^\sigma\tau) \Big((\dot q^\mu- z^\mu)(\dot q^\nu- z^\nu)
+a^\mu a^\nu +b^\mu c^\nu\Big)~.  \nonumber
\end{eqnarray}
Higher order terms are not reported because we
compute the transition amplitude to order $\beta$, for which only the
previous two terms are needed; all fields, classical and
quantum, count as $\beta^{1/2}$.

The transition amplitude can now be computed perturbatively
using Wick-contractions of the quantum fields
\begin{eqnarray}
  \langle x\,\bar\lambda;\beta|y\,\eta;0\rangle &=& A\
  e^{- \tilde S_2}
  \left \langle e^{-S_{int}} \right \rangle\nonumber\\
  &=& A\
  e^{-\frac1{2\beta}g_{\mu\nu} z^\mu z^\nu +\bar\lambda\cdot\eta}
  \exp\Big[- \langle S_3 \rangle-\langle S_4\rangle+\frac12\langle
    S^2_3\rangle_c \Big]
\end{eqnarray}
with the suffix $c$ indicating connected diagrams only and with
the prefactor $A$ soon to be commented upon.
The various Wick-contractions produce
\begin{align} \label{TA-big-expansion}
{}&\exp\Big[-\langle S_3\rangle-\langle S_4\rangle+\frac12\langle
    S^2_3\rangle_c \Big] = \exp\Biggl[-\frac1{4\beta}\partial_\lambda
  g_{\mu\nu}  z^\lambda  z^\mu  z^\nu+\frac1{2\beta}\omega_{\lambda
    ab} z^\lambda\tilde\psi^a\cdot\tilde\psi^b \nonumber\\
{}&- z^\lambda\left(\frac12\partial_\lambda g\, {\bf I}_1+g_\lambda\, {\bf I}_2\right)
-\frac1{12\beta}\partial_\lambda\partial_\sigma g_{\mu\nu}  z^\lambda
 z^\sigma  z^\mu  z^\nu-\frac{\beta}4\partial^2g\, {\bf I}_3-\frac{\beta}2
\partial^\lambda\partial^\sigma g_{\lambda\sigma}\, {\bf I}_4\nonumber\\
{}&+\frac14 \partial_\lambda\partial_\sigma g  z^\lambda  z^\sigma\, {\bf I}_5
+\frac14 \partial^2 g_{\lambda\sigma}  z^\lambda  z^\sigma\, {\bf I}_6
+\partial_\lambda g_\sigma  z^\lambda  z^\sigma\, {\bf I}_7
+\frac1{4\beta}\partial_\lambda \omega_{\sigma ab}
\tilde\psi^a\cdot\tilde\psi^b  z^\lambda z^\sigma\nonumber\\
\displaybreak  %%%%%%%
{}&+\frac1{2}\partial^\lambda \omega_{\lambda ab}
\tilde\psi^a\cdot\tilde\psi^b\, {\bf I}_2 -\frac{\alpha}{\beta}R_{abcd}
\tilde\psi^a\cdot\tilde\psi^b \tilde\psi^c\cdot\tilde\psi^d-\beta V\nonumber\\
{}&-\frac{\beta}4(\partial_\lambda g_{\mu\nu})^2\,
{\bf I}_9-\frac{\beta}2 \partial_\lambda g_{\mu\nu}\partial_\mu
g_{\lambda\nu}\, {\bf I}_{10}-\frac{\beta}8 (\partial_\lambda g)^2\,
{\bf I}_{11} -\frac{\beta}2 g_\lambda \partial_\lambda g\,
{\bf I}_{12} -\frac{\beta}2 g_\lambda^2\, {\bf I}_{13}\nonumber\\
{}&+\frac12 \partial_\lambda
g_{\mu\nu}\partial_\lambda g_{\mu'\nu} z^\mu z^{\mu'}\, {\bf I}_{14}+\frac12 \partial_\lambda
g_{\mu\nu}\partial_\nu g_{\mu'\lambda} z^\mu z^{\mu'}\, {\bf I}_{15}+\frac14 \partial_\lambda
g_{\mu\nu}\partial_{\lambda'} g_{\mu\nu} z^\lambda z^{\lambda'}\, {\bf
  I}_{16}\nonumber\\
{}&+\partial_\lambda
g_{\mu\nu}\partial_{\mu} g_{\mu'\nu} z^\lambda z^{\mu'}\, {\bf I}_{17}+\frac14 \partial_\lambda
g_{\mu\nu}\partial_{\lambda} g  z^\mu z^{\nu}\, {\bf I}_{18}   +\frac12 \partial_\lambda
g_{\mu\nu} g_{\lambda}  z^\mu z^{\nu}\, {\bf I}_{19}\nonumber\\
{}&   +\frac12 \partial_\lambda
g_{\mu\nu} \partial_\mu g  z^\lambda z^{\nu}\, {\bf I}_{20}+\partial_\lambda
g_{\mu\nu}  g_\mu  z^\lambda z^{\nu}\, {\bf
  I}_{21}-\frac1{8\beta}\partial_\lambda g_{\mu\nu} \partial_\lambda
g_{\mu'\nu'} z^\mu z^\nu z^{\mu'} z^{\nu'}\, {\bf I}_{22}\nonumber\\
  {}& -\frac1{2\beta}\partial_\lambda g_{\mu\nu} \partial_{\lambda'}
g_{\lambda\nu'} z^\mu z^\nu z^{\lambda'} z^{\nu'}\, {\bf
  I}_{23}-\frac1{2\beta}\partial_\lambda g_{\mu\nu} \partial_{\lambda'}
g_{\mu\nu'} z^\lambda z^\nu z^{\lambda'} z^{\nu'}\, {\bf
  I}_{24}\nonumber
  \\
{}&+\frac{\beta N}{4} \omega_{\lambda ab}^2\, {\bf I}_{25}
-\frac{N}{4}\omega_{\lambda ab}\,\omega_{\sigma}{}^{ab}  z^\lambda z^\sigma\,
{\bf I}_{26}+\frac{1}{2}\omega_{\lambda ac}\,\omega_{\lambda b}{}^c \tilde\psi^a\cdot\tilde\psi^b\,
{\bf I}_{27}\nonumber\\
{}&-\frac{1}{2\beta}\omega_{\lambda ac}\,\omega_{\sigma b}{}^c
 z^\lambda  z^\sigma \tilde\psi^a\cdot\tilde\psi^b\,
{\bf I}_{28}-\frac{1}{8\beta}\omega_{\lambda ab}\,\omega_{\lambda cd}
\tilde\psi^a\cdot\tilde\psi^b\tilde\psi^c\cdot\tilde\psi^d\,{\bf
  I}_{29}\nonumber\\
{}&+\frac14 \omega_{\lambda ab}\partial_\lambda
g\tilde\psi^a\cdot\tilde\psi^b\,{\bf I}_{30}+\frac12 \omega_{\lambda ab}
g_\lambda \tilde\psi^a\cdot\tilde\psi^b\,{\bf I}_{31}-\frac1{4\beta} \omega_{\lambda ab}
\partial_\lambda g_{\mu\nu}  z^\mu  z^\nu \tilde\psi^a\cdot\tilde\psi^b\,{\bf I}_{32}
\nonumber\\{}&-\frac1{2\beta} \omega_{\mu ab}
\partial_\lambda g_{\mu\nu}  z^\lambda  z^\nu \tilde\psi^a\cdot\tilde\psi^b\,{\bf I}_{33}
\Biggr]
\end{align}
where we made use of several shortcut notations, including $g_\mu =
g^{\lambda\nu}\partial_\lambda g_{\mu\nu}$, $\partial_\lambda g
=g^{\mu\nu}\partial_\lambda g_{\mu\nu}$, $\partial_\lambda
\partial_\sigma g =g^{\mu\nu}\partial_\lambda \partial_\sigma
g_{\mu\nu}$ and $\partial^2
=g^{\lambda\sigma}\partial_\lambda\partial_\sigma$. The integrals ${\bf
  I}_k$ are reported in appendix~\ref{app:FD} along with the
pictorial representation of the diagrams they belong to
(an integral named ${\bf I}_8$ is absent, but we kept the same notation as in \cite{Bastianelli:2006rx},
where such an integral arose from the coupling to external gauge fields,
to allow easy comparison).
We compute them in the following subsections,
using the different regularization schemes discussed in the introduction.
The purely bosonic contributions ($k\leq 24$) are well-known from many previous
computations, see \cite{Bastianelli:2006rx} for example; the remaining ones have
been computed, though only for $N\leq 2$, in the time slicing regularization technique \cite{deBoer:1995cb},
in dimensional regularization \cite{Bastianelli:2002qw,Bastianelli:2005vk},
and in mode regularization \cite{Bonezzi:2008gs}. However, in the last two schemes
fermions were traced out to obtain directly heat kernel coefficients and
trace anomalies. In the present case we  are interested
in the full transition amplitude and wish to keep $N$ generic, so that we need to
reconsider all such integrals with fermionic contributions.
Finally, the prefactor $A$ can be fixed
by requiring that, in the limit $\beta \to 0$, the transition amplitude
reduces to  $ \langle x\, \bar\lambda;\beta|y\, \eta;0\rangle \to
\delta(x-y)\, e^{\bar\lambda\cdot \eta}$, which in MR and DR gives
\begin{equation}
  A = \frac1{(2\pi\beta)^{D/2}}~.
\end{equation}
In TS such prefactor can be deduced directly starting from operatorial methods, and reads
\begin{equation} \label{TS-prefactor}
  A = \left[\frac{g(x)}{g(y)}\right]^{1/4}\frac1{(2\pi\beta)^{D/2}}~.
\end{equation}
We are now ready to consider the various regularization schemes.

%%%%%%%%%%%%%%%%%%%%%%%%%%%%%%%%%%%%%%%%
\subsection{Time slicing regularization}
\label{subs:TS}

Time sliced path integrals in curved space were extensively discussed in~\cite{DeBoer:1995hv,deBoer:1995cb}.
In essence time slicing regularization consists in studying the discretized version of the path integral,
 as derived form the operatorial picture by using Weyl ordering and midpoint rule,
 and recognizing the action with the correct counterterms together with the rules how to compute Feynman
graphs that must be used in the continuum limit. These rules state that the Dirac delta functions should
always be implemented as if they were Kronecker delta's,
using the value $\theta(0) = \frac12$ for the discontinuous step function.
We do not need to repeat here many computations, namely the bosonic ones,
though one can easily compute them using the explicit expressions collected in appendix~\ref{app:FD},
or extract them directly from \cite{Bastianelli:2006rx}.
It is enough to focus on the graphs containing fermionic lines, i.e.  ${\bf I}_k$ with $k\geq 25$.

Thus, let us start with ${\bf I}_{25}$, that is the only diagram that depends heavily 
on the regularization chosen. Following the TS prescriptions, we have
\begin{eqnarray}
{\bf I}_{25}&=& \int_{-1}^0\int_{-1}^0d\tau d\sigma
   ~\ddeld~\del_F^2
   =\int_{-1}^0\int_{-1}^0d\tau d\sigma
   \Big (1- \delta(\tau-\sigma) \Big ) \Big(\frac12 \epsilon (\tau-\sigma) \Big )^2 \nonumber \\
   &=&\frac14 \int_{-1}^0\int_{-1}^0d\tau d\sigma\,  \epsilon^2 (\tau-\sigma)  =
   \frac14 \int_{-1}^0\int_{-1}^0d\tau d\sigma   =\frac14
\end{eqnarray}
since $\ddeld=1- \delta(\tau-\sigma)$ and $\epsilon (0)=0$, as follows form $\theta (0)=\frac12$
and eq. (\ref{eq:Df}).
The regular ${\bf I}_{26}$ does not need any prescription
and corresponds to the last line above, giving ${\bf I}_{26}=\frac14$.
Similarly, one finds that the integrals ${\bf I}_k$ with $k\geq 27$ yield a vanishing result
(those with $k\geq 28$ actually contain only bosonic propagators, but depend on the external fermionic backgrounds).

The transition amplitude then reads
\begin{eqnarray}
  &&\langle x\,\bar\lambda;\beta|y\,\eta;0\rangle =
\frac{e^{-\frac1{2\beta}g_{\mu\nu} z^\mu z^\nu
    +\bar\lambda\cdot\eta}}{(2\pi\beta)^{D/2}}
\exp\Big[\langle {\rm bosonic}\rangle_{_{TS}}
+\frac1{2\beta} \omega_{\lambda ab} z^\lambda
    \tilde\psi^a\cdot\tilde\psi^b\nonumber\\&&
+\frac1{4\beta} \partial_\lambda\omega_{\sigma
      ab} z^\lambda  z^\sigma \tilde\psi^a\cdot\tilde\psi^b
%    \nonumber\\&&
-\frac{\alpha}{\beta}
    R_{abcd}\tilde\psi^a\cdot\tilde\psi^b\tilde\psi^c\cdot\tilde\psi^d
-\frac{N}{16} \omega_{\lambda ab}\omega_{\sigma ab}  z^\lambda
 z^\sigma\nonumber\\
&&+\frac{\beta N}{16} \omega_{\lambda ab}^2\Big]
\label{Tamp-TS}
\end{eqnarray}
where $\langle {\rm bosonic}\rangle_{_{TS}}$ contains the purely bosonic contributions of
\eqref{TA-big-expansion},
including the metric factors appearing in \eqref{TS-prefactor}. It can be extracted from the literature, or easily
recomputed with the set-up described above,
and reads
\begin{align}
\langle {\rm bosonic}\rangle_{_{TS}} =
&-\frac{1}{4\beta}\,\de_\mu g_{\nu\lambda}\,z^\mu z^\nu z^\lambda-\frac{1}{12\beta}\,\Big(\de_\mu\de_\nu g_{\lambda\sigma}-\frac12\,g_{\rho\tau}\,\Gamma^\rho_{\mu\nu}\,\Gamma^\tau_{\lambda\sigma}\Big)\,z^\mu z^\nu z^\lambda z^\sigma \nonumber\\ 
& +\frac{1}{12}R_{\mu\nu}z^\mu z^\nu+\beta\Big(
\frac18 g^{\mu\nu}\Gamma_{\mu\lambda}^{\sigma} \Gamma_{\nu\sigma}^{\lambda}
-\frac{1}{24}R  - V \Big)\;.
\end{align}

This transition amplitude is the one computed with a TS regularization of the  Feynman
diagrams, and in general differs form those computed with other regularizations if no counterterms
are introduced.  In particular,  eq. \eqref{Tamp-TS} satisfies a Schr\"odinger equation
with a non-covariant hamiltonian that differs from the one given in  eq. \eqref{eq:H}.
One expects different regularizations to be related by finite local counterterms, so we need
to identify the correct counterterm that make sure that we are discussing the same quantum theory
as the one associated to the hamiltonian $H$ of eq. \eqref{eq:H}.
To achieve this we can either compare with the transition amplitude obtained by operatorial methods,
or compute directly the hamiltonian associated with the transition amplitude
above. We shall follow both methods as a check of our calculations.

As it stands the transition amplitude calculated above cannot be compared directly with the result from
canonical methods in  eq. \eqref{final result TA}:  the latter is valid for even $N$
and is written in terms of fermionic coherent states that correspond to the physical fermions only,
while the present result, valid for any integer $N$, contains also the Majorana spectator fields introduced 
in the fermion doubling trick.
To overcome these differences, we can take a trace over the fermionic Hilbert space and eliminate the contribution
of the decoupled spectator variables by subtracting  their degrees of freedom.
 Thus, let us compute the trace in the fermionic sector of eq. \eqref{Tamp-TS} by integrating
 over the Grassmann variables  with measure $ \int d\eta d\bar\lambda \, e^{\bar\lambda \cdot \eta} $,
 see appendix  \ref{app:FCS}.
 The final result can be obtained, for instance, using the standard Wick-contractions associated to the gaussian integral
 $ \int d\eta d\bar\lambda \, e^{2\bar\lambda \cdot \eta}$
(note the factor 2 in the exponent arising form the trace measure and the leading part of \eqref{Tamp-TS}),
which gives the following propagators
\begin{equation}
 \langle \bar \lambda^a_i\eta^b_j\rangle= \frac12 \eta^{ab} \delta_{ij}  \; \quad \to \quad
  \langle \tilde\psi^a_i \tilde\psi^b_j \rangle= 0
   \end{equation}
where we used the definition in \eqref{tilde fermion}. Note that the normalization factor
 $ \int d\eta d\bar\lambda \, e^{2\bar\lambda \cdot \eta} = 2^{ND}$
has to be renormalized to $2^{\frac{ND}{2}}$ to undo the fermion doubling.
Taking all this into account, and setting $z\equiv y-x=0$ for simplicity,
we obtain
\begin{equation}  2^{-\frac{ND}{2}} \!\!
\int d\eta d\bar\lambda \ e^{\bar\lambda \cdot \eta}
\langle x\,\bar\lambda;\beta|x\,\eta;0\rangle =
\frac{2^\frac{N D}{2}}{(2\pi\beta)^{D/2}}\,
\Big[1+
\beta\Big( \frac18 g^{\mu\nu}\Gamma_{\mu\lambda}^{\sigma} \Gamma_{\nu\sigma}^{\lambda}
-\frac{1}{24}R  - V \Big ) +\frac{\beta N}{16} \omega_{\lambda ab}^2
+... %\mathcal{O}(\beta^2)
\Big]
   \end{equation}
where the first term in round brackets is due to the purely bosonic contributions.
Comparing with \eqref{traced TA-coinciding points} one recognizes the counterterm
that needs to be added
to the potential $V$ in the path integral action to achieve equality
\begin{eqnarray} \label{TS counterterm}
V_{TS}^{(N)}&=&-\Big(\frac18+\frac{\alpha N}{2}\Big) R
+\frac18 g^{\mu\nu}\Gamma_{\mu\lambda}^{\sigma} \Gamma_{\nu\sigma}^{\lambda}
+\frac{N}{16}\omega_{\mu ab}\omega^{\mu ab}~.
\label{eq:VTS}
\end{eqnarray}
Power counting shows that no higher order corrections to the counterterm are to be expected.

Alternatively, one may compute the hamiltonian appearing
in the Schr\"odinger equation satisfied by
the amplitude \eqref{Tamp-TS}. To do this we  insert the latter
into~(\ref{eq:evolution}), Taylor-expand to order $\beta$ all terms in
the right hand side about the final point,
and identify the Schr\"odinger equation associated to it.
Comparing with the Schr\"odinger equation due to the hamiltonian
(\ref{eq:H}) one deduces eventual counterterms. Let us describe this computation.
We perform the Gaussian integrals over $d^D y= d^D z$ and the integrals over the fermionic coherent states
using the properties summarized in appendix~\ref{app:FCS}. The purely bosonic
contributions of the diagrammatic expansion yield the standard TS result
\begin{eqnarray}
\Phi(x,\bar\lambda;\beta) &=& \Big(1-\beta\partial_t -\beta
H_B+O(\beta^2)\Big)
\Phi(x,\bar\lambda;\beta)
\label{eq:phi}
\end{eqnarray}
with
\begin{eqnarray}
H_B =-\frac1{2\sqrt{g}}\partial_\mu
g^{\mu\nu}\sqrt{g}\partial_\nu+V+\frac18
\Big(R- g^{\mu\nu}\Gamma_{\mu\lambda}^{\sigma} \Gamma_{\nu\sigma}^{\lambda} \Big)~.
\end{eqnarray}
that by itself requires the addition of the familiar counterterm $V^{(0)}_{TS}
=-\frac18 \Big(R- g^{\mu\nu}\Gamma_{\mu\lambda}^{\sigma} \Gamma_{\nu\sigma}^{\lambda}\Big)$
into the path integral, in order to get the covariant
 $H = -\frac1{2\sqrt{g}}\partial_\mu g^{\mu\nu}\sqrt{g}\partial_\nu +V$.
 We need to dress this result with the fermionic contributions.
 One way to perform the integrals over the fermionic coherent states
 is to use the formulas involving fermionic delta functions derived at 
  the end of section~\ref{app:FCS}. By defining the full hamiltonian $H_F$ as
\begin{eqnarray}
H_F =H_B+\Delta
\end{eqnarray}
and using~\eqref{eq:phi}, we obtain
\begin{eqnarray}
\Delta &=&-\frac14 \partial^\lambda \omega_{\lambda ab} M^{ab}
  +\frac14 g^{\lambda\sigma} \Gamma^\mu_{\lambda\sigma} \omega_{\mu ab}
  M^{ab} -\frac12 \omega^\mu{}_{ab} M^{ab}
  \partial_\mu -\frac18 \omega^\lambda{}_{ab}\omega_{\lambda cd}
  M^{ab} M^{cd}\nonumber\\&&+\alpha R_{abcd} M^{ab}
  M^{cd}+\frac{\alpha N}{2} R-\frac{N}{16} (\omega_{\lambda ab})^2
\label{eq:delta-TS'}
\end{eqnarray}
with $M^{ab}=
\frac12\Big(\bar\lambda^a\cdot\bar\lambda^b+\bar\lambda^a\cdot\frac{\partial}{\partial\bar
 \lambda_b}-\bar\lambda^b\cdot\frac{\partial}{\partial\bar
 \lambda_a} +\frac{\partial}{\partial\bar
 \lambda_a}\cdot\frac{\partial}{\partial\bar
 \lambda_b}\Big)
$ being Lorentz generators.
We observe that the noncovariant terms in the line above are those
necessary to render the bosonic hamiltonian $H_B$ local-Lorentz
covariant, namely
\begin{eqnarray}
H_F &=&H_B+\Delta = -\frac12 g^{\mu\nu} D_\mu D_\nu +\alpha R_{abcd}
M^{ab} M^{cd} +V\nonumber\\
&&+\Big(\frac18+\frac{\alpha N}{2}\Big)
R-\frac18g^{\lambda\sigma}\Gamma_{\lambda\rho}^{\tau}\Gamma_{\sigma\tau}^{\rho}
-\frac{N}{16} (\omega_{\lambda ab})^2
\end{eqnarray}
where $D_\mu$ is the fully covariant derivative (with both Lorentz and Christoffel connections),
so that in order to have\footnote{Here we use $\langle \bar\lambda| \psi^{ai}|\Phi\rangle =
\frac1{\sqrt2}\left(\frac{\partial}{\partial \bar \lambda^i_a}+\bar
\lambda^{ai}\right)\Phi(\bar\lambda)$, so that the Lorentz
generators can be written as $M^{ab} =\frac12\left(
\psi^a\cdot\psi^b-\psi^b\cdot\psi^a\right)$.}
\begin{eqnarray}
H % &=& -\frac12 g^{\mu\nu} D_\mu D_\nu +\alpha R_{abcd}
% M^{ab} M^{cd}+V\nonumber\\
&=& -\frac12 g^{\mu\nu} D_\mu D_\nu +\alpha R_{abcd}
\psi^a\cdot\psi^b\psi^c\cdot\psi^d +V
\label{qH}
\end{eqnarray}
one needs to add to the path integral the same counterterm found before in eq. \eqref{TS counterterm}.
Thus we found complete agreement.

The above expression for the counterterm  $V_{TS}^{(N)}$ matches all the previously known
cases~\cite{DeBoer:1995hv,deBoer:1995cb}: the
purely bosonic case ($N=0$) is  obviously reproduced.
For $N=1$ supersymmetry fixes $\alpha= -1/4$ and $V=0$ so that
$V_{TS}^{(1)}=\frac18 g^{\mu\nu}\Gamma_{\mu\lambda}^{\sigma} \Gamma_{\nu\sigma}^{\lambda}
+\frac{1}{16}(\omega_{\mu ab})^2$ comes out correctly. Note that in this case the relation
$R_{abcd} \psi^a\psi^b\psi^c\psi^d=-\frac12 R$ allows to use $\alpha =0$ and  $V=\frac18 R$
as well to identify the same supersymmetric hamitonian.
For $N=2$, supersymmetry requires $\alpha=-\frac18$ and $V=0$, so that again
$V_{TS}^{(2)}=\frac18 g^{\mu\nu}\Gamma_{\mu\lambda}^{\sigma} \Gamma_{\nu\sigma}^{\lambda}
+\frac{1}{8}(\omega_{\mu ab})^2$ is correctly reproduced.

%%%%%%%%%%%%%%%%%%%%%%%%%%%%%%%%%%%%%%%%%
\subsection{Mode regularization}
\label{subs:MR}
In this section we approach the previous computation using mode regularization (MR), that can be considered 
as the worldline equivalent of a cut-off regularization in momentum space of standard quantum field theories.
Mode regularization starts by expanding all fields in Fourier sums, thus identifying a suitable functional space to
integrate over in the path integral.
The regularization is achieved by truncating the Fourier sums at a fixed mode $M$, so that all distributions  
appearing in Feynman graphs become well-behaved functions. Eventually 
one takes the limit $M \to \infty$ to obtain a unique (finite) result. 
Often one can proceed faster: performing manipulations that are guaranteed to be valid at the regulated level 
one may cast the integrands in alternative forms that can be computed directly in the continuum limit,
without encountering any ambiguity.

The boundary conditions for the bosonic variables are as in (\ref{bc-uno}), 
so that the bosonic quantum fluctuations, as well as the ghosts, are naturally expanded in a Fourier sine series 
\begin{equation}
q^\mu(\tau)=\sum_{m=1}^M q^\mu_m\sin(\pi m\tau)
\end{equation}
where the mode $M$ is the regulating cut-off that is eventually sent to infinity,  as in 
\cite{Bastianelli:1991be}. This choice preserves the boundary conditions imposed at initial and final times.
On the other hand, fermionic fields satisfy first order differential equation and carry boundary conditions only at initial or final times, 
but not at both, see eqs. (\ref{eq:P}) and (\ref{eq:barP}). Thus we find it useful to expand the fermionic quantum fields
in half integers modes as follows
\begin{equation} 
{\bar Q}^{a i}(\tau)=\sum_{r=\frac12}^{M-\frac12} \bar Q^{a i}_{r}\sin(\pi r \tau)\;, \qquad
Q^{a i}(\tau)=\sum_{r=\frac12}^{M -\frac12}Q^{a i}_r \cos(\pi r\tau) \;.
\end{equation} 
This choice preserves the boundary conditions and provides us with  
a regulated functional space to integrate over also for the fermions.
In addition, it produces a simple regulated kinetic term that is easily inverted to obtain the propagators.
Finally, the path integral is defined as a regulated integration over the
Fourier coefficients  of the various fields.

Perturbatively, the propagators are as in eqs. (\ref{qq})--(\ref{eq:QQ})
and are regulated  as follows 
\begin{eqnarray}
  \Delta(\tau,\sigma) &=&-\sum_{m=1}^{M}\frac{2}{\pi^2m^2}\sin{(\pi
m\tau)}\sin{(\pi m\sigma)}\\
\Delta_{gh}(\tau,\sigma) &=&\sum_{m=1}^{M} 2 \sin{(\pi m\tau)}\sin{(\pi m\sigma)}\\
 G(\tau,\sigma)&=&\sum_{r=\frac1 2}^{M-\frac 1 2} \frac{2}{\pi r} \sin{(\pi r\tau)}\cos{(\pi r\sigma)} \;.
\end{eqnarray}
As a consequence
\begin{equation}
  \Delta_F(\tau,\sigma) =\sum_{r=\frac12}^{M-\frac12}
  \frac{1}{\pi r}\sin{(\pi r (\tau-\sigma))}
\end{equation}
for (\ref{chichi}) that turns out to be translational invariant.
Note that in the limit $M\rightarrow\infty$ the previous formulas
reproduce eqs.~\eqref{eq:D} and~\eqref{eq:Df}.

We are now ready to compute the Feynman integrals  with fermionic contributions in MR 
(the purely bosonic ones are standard and can be extracted from \cite{Bastianelli:2006rx}).
Using the regularized expressions one obtains  
$\mathbf{I}_{27}=\mathbf{I}_{28}=...=\mathbf{I}_{33}=0$.
Also, one finds $\mathbf{I}_{26}=\frac14$, as it is unambiguous and gives the same result in all regularization schemes.
The only tricky integral is $\mathbf{I}_{25}$. 
Using the mode regulated propagators we can manipulate it 
using an integration by parts as follows
\begin{equation}
\begin{split}
\mathbf{I}_{25} &= 
\int_{-1}^0\int_{-1}^0d\tau d\sigma\,\ddeld\,\Delta_F^2= 
\int_{-1}^0 d\tau \Big [\deld \Delta_F^2 \Big]_{\sigma=-1}^{\sigma=0}
- \int_{-1}^0\int_{-1}^0d\tau d\sigma\,\deldd\,\Delta_F^2= 
 \\
&=
\frac12 \int_{-1}^0 d\tau \Big [\deld \Delta_F^2 \Big]_{\sigma=-1}^{\sigma=0}
+\frac12 \int_{-1}^0\int_{-1}^0d\tau d\sigma\,( \ddeld -\deldd)\,\Delta_F^2
\end{split}
\end{equation}
The first term is unambiguous and
can be computed directly upon removing the cut-off to give $\frac18$. 
In the second term one can use the following identity valid at the regulated level (with $x=\tau-\sigma$)
\begin{equation}
\ddeld(\tau,\sigma) -\deldd(\tau,\sigma)  = -2 \cos\Big (\frac{\pi x}{2}\Big)\, \ddel_F(x) +1 -\cos(\pi M x) \;.
\end{equation}
The integral involving the last term ($\cos(\pi M x)$)
produces a vanishing result in the limit $M\to \infty$, thanks to the Riemann-Lebesgue lemma,
while the rest can be computed using an integration by parts and removing the regularization
to yield the final result
\begin{equation}
\mathbf{I}_{25} = \frac18 + \int_{-1}^0\int_{-1}^0d\tau d\sigma\, \Big [
\frac12 -\cos\Big (\frac{\pi x}{2}\Big)\, \ddel_F(x) \Big ]
\Delta_F^2 (x)=
\frac16 \;.
\end{equation}

The transition amplitude computed in mode regularization hence reads
\begin{eqnarray}
  &&\langle x\,\bar\lambda;\beta|y\,\eta;0\rangle =
\frac{e^{-\frac1{2\beta}g_{\mu\nu}z^\mu z^\nu
    +\bar\lambda\cdot\eta}}{(2\pi\beta)^{D/2}}
\exp\Big[\langle {\rm bosonic}\rangle_{_{MR}}
+\frac1{2\beta} \omega_{\mu ab}\,z^\mu
    \,\tilde\psi^a\cdot\tilde\psi^b\nonumber\\
    &&+\frac1{4\beta}\, \partial_\mu\omega_{\nu ab}\,z^\mu z^\nu\,
\tilde\psi^a\cdot\tilde\psi^b
-\frac{\alpha}{\beta}\,R_{abcd}\,\tilde\psi^a\cdot\tilde\psi^b\tilde\psi^c\cdot\tilde\psi^d
-\frac{N}{16}\, \omega_{\mu ab}\,\omega_{\nu}{}^{ab}\, z^\mu z^\nu\nonumber\\
&&+\frac{\beta N}{24}\, \omega_{\mu ab}^2\Big]~.
\label{Tamp-MR}
\end{eqnarray}
where $\langle {\rm bosonic}\rangle_{_{MR}}$ is the purely bosonic contribution that can be found in
\cite{Bastianelli:2006rx}.
Comparing the above result with the TS one given in eq. \eqref{Tamp-TS}, we notice that the only
difference coming from the fermionic sector sits in the coefficients of the last $\omega^2$
term, which is due to $\mathbf{I}_{25}$. As the bosonic part of the MR calculation requires the counterterm
$$ V^{(0)}_{MR} =-\frac18 R-\frac{1}{24}\left(\Gamma^\mu_{\nu\lambda}\right)^2 $$
to reproduce the heat kernel for $H=-\frac{1}{2\sqrt{g}}\de_\mu g^{\mu\nu}\sqrt{g}\de_\nu$,  
where $\left(\Gamma^\mu_{\nu\lambda}\right)^2\equiv
g^{\nu\nu'}g^{\lambda\lambda'}g_{\mu\mu'}\Gamma^\mu_{\nu\lambda}\Gamma^{\mu'}_{\nu'\lambda'}$,
we see that, in order to obtain the correct amplitude
for \eqref{eq:H}, we need the counterterm
\begin{equation}
V_{\rm{MR}}^{(N)}=-\Big(\frac18+\frac{\alpha N}{2}\Big) R
-\frac{1}{24}\left(\Gamma^{\mu}_{\nu\lambda}\right)^2
+\frac{N}{24}\omega_{\mu ab}\omega^{\mu ab}
\end{equation}
that again matches the results known for $N=0,1,2$
\cite{Bastianelli:1998jm,Bonezzi:2008gs}.

%%%%%%%%%%%%%%%%%%%%%%%%%%%%%%%%%%%%%%%
\subsection{Dimensional regularization}
\label{subs:DR}

Finally, we reconsider the previous calculations using  dimensional regularization. 
This is a perturbative regularization that
uses an adaptation of standard dimensional regularization to regulate the distributions defined on the 
one dimensional compact space  $I=[0,\beta]$. 
One adds $d$ extra non-compact dimensions and perform all computations 
of ambiguous Feynman graphs in $d + 1$ dimensions. 
Extra dimensions act as a regulator when $d$ is extended analytically to the complex plane, as in the usual QFT case.
After evaluation of the various integrals one should take the $d\to 0$ limit  \cite{Bastianelli:2000nm}.
This is quite difficult in general, since the compact space $I$ produces sums over discrete momenta, and  
standard formulas of dimensional regularization do not include that situation. 
However one may use manipulations valid at the regulated level, 
like differential equations satisfied by the Green functions and partial integration, 
to cast the integrands in equivalent forms that
can be unambiguously computed in the  $d\to 0$  limit. 
While purely perturbative, this method carries a covariant counterterm
that simplify extensive calculations, as the one performed in \cite{Bastianelli:2000dw}
to obtain trace anomalies  for a scalar field in six dimensions using worldline methods.

In DR the computation turns out to be quite simple for most
diagrams: ${\bf I}_{27}={\bf I}_{28}=0$ because the integrand is odd,
whereas ${\bf I}_{29}=\cdots={\bf I}_{33}=0$ as can be  shown by
integrating by parts. Also ${\bf I}_{26}=\frac14,$ as it is regular and
can be safely evaluated by using the unregulated expression for the
propagator. As usual, more care is needed to compute ${\bf I}_{25}$ since the
integral is ambiguous (products of distributions). By dimensionally
extending the cubic vertex
\begin{eqnarray}
\dot x^\mu \psi^a_i\psi^b_i\quad\to\quad \partial_A x^\mu
\bar {\psi}^a_i \gamma^A\psi^b_i\,,\quad A=1,\dots,d+1
\end{eqnarray}
the above integral becomes
\begin{eqnarray}
{\bf I}_{25}&=&\int_{-1}^0\int_{-1}^0d\tau d\sigma
   ~\ddeld~\del_F^2 \quad \to \quad \int\int dt dt'~ {}_A\del_{B} ~{\rm tr} 
   [\gamma^A\del_F\gamma^B\del_F]
\end{eqnarray}
where $\del(t,t')=-\beta^{-1}\langle q(t) q(t')\rangle$ and
$\del_F(t,t') =\beta^{-1}\langle \chi(t) \bar\chi(t')\rangle$
(the bar indicates Majorana
conjugation, see \cite{Bastianelli:2000nm})
are the dimensionally regulated propagators
where $ \partial \!\!\! / \del_F
=-\del_F \!\!\stackrel{\gets}{\partial \!\!\! / '}
= \delta^{(d+1)}(t,t')$, and $ {}_A\del_{B}(t,t')=\partial_A\partial'_B \del(t,t')$.
One can partially
integrate the above integral without picking boundary terms since in the compact
direction bosonic fields vanish at the boundary, and in the extended
directions Poincar\'e invariance allows partial integration as usual. Hence
the above expression becomes
\begin{eqnarray}
&&-\int\int dt dt'~ \del_B ~\partial_A {\rm tr} [
\gamma^A\del_F \gamma^B\del_F] \nonumber\\
&&=-\int\int dt dt'~ \del_B ~{\rm tr}
\left[ (\partial \!\!\! /  \del_F) \gamma^B \del_F -
\del_F \gamma^B (\del_F  \!\!\stackrel{\gets}{\partial \!\!\! / '})\right]
\nonumber\\
&&=-2\int dt ~\del_B ~{\rm tr}
[\gamma^B\del_F(t,t)] \quad\to\quad
{\bf I}_{25}=2\int_{-1}^0d\tau~ \ddel~\del_F|_\tau =0
\label{I25DR}
\end{eqnarray}
where we have used the regulated Green equation, integrated the delta function, and used that $\del_F$
vanishes for coinciding points.

 The transition amplitude in DR thus reads
\begin{eqnarray}
  &&\langle x\,\bar\lambda;\beta|y\,\eta;0\rangle =
\frac{e^{-\frac1{2\beta}g_{\mu\nu} z^\mu z^\nu
    +\bar\lambda\cdot\eta}}{(2\pi\beta)^{D/2}}
\exp\Big[\langle {\rm bosonic}\rangle_{_{DR}}
+\frac1{2\beta} \omega_{\lambda ab} z^\lambda
    \tilde\psi^a\cdot\tilde\psi^b\nonumber\\&&
+\frac1{4\beta} \partial_\lambda\omega_{\sigma
      ab} z^\lambda  z^\sigma \tilde\psi^a\cdot\tilde\psi^b
%    \nonumber\\&&
-\frac{\alpha}{\beta}
    R_{abcd}\tilde\psi^a\cdot\tilde\psi^b\tilde\psi^c\cdot\tilde\psi^d
-\frac{N}{16} \omega_{\lambda ab}\omega_{\sigma ab}  z^\lambda  z^\sigma\Big]~.
\end{eqnarray}
where the bosonic contribution $\langle {\rm bosonic}\rangle_{_{DR}}$ can be extracted 
from \cite{Bastianelli:2006rx}.
Comparing this result with the ones obtained before in other regularizations 
we note that, thanks to the vanishing of ${\bf I}_{25}$, 
the diagrammatic expansion does not produce the term proportional to $\omega^2$
which previously had to be canceled by the counterterms.
Thus,  the standard bosonic counterterm  $V_{DR}^{(0)}=-\frac18 R$ is dressed up to 
\begin{eqnarray}
V_{DR}^{(N)}=-\Big(\frac18+\frac{\alpha N}{2}\Big) R~.
\end{eqnarray}
This  matches known results for $N=0,1,2$
\cite{Bastianelli:2000nm, Bastianelli:2002qw, Bastianelli:2005vk}.

\section{Conclusions}
In this paper we have studied the quantum mechanics of one dimensional nonlinear sigma models 
possessing a O($N$) extended supersymmetry on suitable target space backgrounds.
Considering an arbitrary background geometry,
we have computed the transition amplitude for short propagation times 
using both canonical and path integrals methods, obtaining in the latter case the 
correct counterterms associated to various regularization schemes which are needed to define
unambiguously the path integrals. 

A possible use of our results is  in the discussion of higher spin fields in a first 
quantized picture. Worldline approaches are useful in finding efficient ways of computing 
amplitudes for relativistic processes both in flat space \cite{Schubert:2001he}
and curved spaces \cite{Bastianelli:2002fv}. The quantum mechanical 
nonlinear sigma models discussed here arose precisely in an attempt to use 
worldlines methods to describe one-loop effective action due to higher spin fields
in a curved background \cite{Bastianelli:2007pv, Bastianelli:2008nm, Corradini:2010ia}.
In future works we  plan indeed to use the path integrals constructed here 
to study effective actions induced by higher spin fields and compute 
corresponding heat kernel coefficients.
Finally, one might wish to extend  our results to the OSp($N,2M$) quantum mechanics of  
ref. \cite{Hallowell:2007qk},
which have also found applications to higher spin theories \cite{Bastianelli:2009eh, Cherney:2009mf, Cherney:2009md}.

\newpage

%%%%%%%%%%%%%%%%%%%%%%%%%%%
\appendix

%%%%%%%%%%%%%%%%%%%%%%%%%%%%%%%%%%%%%
\section{Fermionic coherent states}
\label{app:FCS}
The even-dimensional Clifford algebra
\begin{eqnarray}
  \{\psi^M,\psi^N\}=\delta^{MN}\,,\quad M,N=1,\dots,2l
\end{eqnarray}
can be written as a set of $l$ fermionic harmonic oscillators (the
index $M$ may collectively denote a set of indices that may involve
internal indices as well as a space-time index), by
simply taking complex combinations of the previous operators
\begin{eqnarray}
  && a^m =\frac1{\sqrt{2}} \left(\psi^{m}+i\psi^{m+l}\right)\\
  && a^\dagger_m =\frac1{\sqrt2}
  \left(\psi^{m}-i\psi^{m+l}\right)\,,\quad\quad m=1,\dots,l\\
  && \{a^m,a^{\dagger}_{n}\} = \delta_{n}^m
\end{eqnarray}
and it can be thus represented in the vector space spanned by the $2^l$ orthonormal states $|{\bf
  k}\rangle = $ $\prod_m (a_m^\dagger)^{k_m} |0\rangle$ with
$a_m|0\rangle =0$ and the vector $\bf k$ has elements taking only two possible values,
$k_m=0,1$. This basis (often called spin-basis)
  yields a standard representation of the Clifford algebra,
  i.e. of the Dirac gamma matrices.

An alternative overcomplete basis is given by the coherent states
that are eigenstates of creation or annihilation operators
\begin{eqnarray}
  && |\xi\rangle= e^{a_m^\dagger \xi^m} |0\rangle\quad\rightarrow\quad
  a^m|\xi\rangle = \xi^m |\xi\rangle = |\xi\rangle \xi^m \\
  && \langle \bar\eta| = \langle 0| e^{\bar \eta_m a^m}  \quad\rightarrow\quad
  \langle \bar\eta|a^\dagger_m = \langle \bar \eta|\bar\eta_m =
  \bar\eta_m \langle \bar \eta| ~.
\end{eqnarray}
Below we list some of the useful properties satisfied by these
states. Using the Baker-Campbell-Hausdorff formula $e^X e^Y = e^Y e^X
e^{[X,Y]}$, valid if $[X,Y] = c$-number, one finds
\begin{eqnarray}
  \langle \bar \eta | \xi\rangle = e^{\bar\eta\cdot \xi}
\end{eqnarray}
that in turn implies
\begin{eqnarray}
  && \langle \bar \eta | a^m | \xi \rangle = \xi^m \langle \bar \eta |
  \xi\rangle= \frac\partial{\partial \bar\eta_m}\langle \bar \eta |
  \xi\rangle\\
  &&\langle \bar \eta | a^\dagger_m | \xi\rangle = \bar\eta_m \langle \bar \eta |
  \xi\rangle
\end{eqnarray}
so that $\{\frac\partial{\partial
  \bar\eta_m},\bar\eta_{n}\}=\delta_{n}^m$.
 Defining
\begin{eqnarray}
  \label{bar-eta}
  d \bar\eta = d\bar\eta_l\cdots d\bar\eta_1\;, \qquad
  d\xi = d\xi^1\cdots d\xi^l
\end{eqnarray}
so that $d\bar\eta d\xi %\equiv \prod_m d\bar\eta_md\xi^m
=d\bar\eta_1 d\xi^1d\bar\eta_2 d\xi^2\cdots d\bar\eta_l d\xi^l$, one finds the following relations
\begin{eqnarray}
  &&\int d\bar\eta d\xi\ e^{-\bar\eta\cdot \xi} =1\label{10}\\
  &&\int d\bar\eta d\xi\ e^{-\bar\eta\cdot \xi}\ |\xi\rangle\langle
  \bar\eta|={\mathbb 1}\label{11}
\end{eqnarray}
where ${\mathbb 1}$ is the identity in the Fock space.
One can also define a fermionic delta function
with respect to the measure (\ref{bar-eta}) by
\begin{eqnarray}
    \delta(\bar\eta-\bar\lambda) \equiv 
(\bar\eta^1-\bar\lambda^1)\cdots(\bar\eta^l-\bar\lambda^l) =  
\int d\xi\ e^{(\bar\lambda-\bar\eta)\cdot\xi}  \;.
\end{eqnarray}
Finally, the trace of an arbitrary operator can be written  as
\begin{eqnarray}
  {\rm Tr}\, A =
    \int d\bar\eta d\xi\ e^{-\bar\eta\cdot \xi}  \langle-\bar\eta| A |\xi\rangle
    = \int     d\xi d\bar\eta
    \ e^{\bar\eta\cdot \xi}  \langle\bar\eta| A |\xi\rangle \;.
\end{eqnarray}
As a check one may compute the trace of the identity
\begin{eqnarray}
  {\rm Tr}\, {\mathbb 1} &=& \int d\xi d\bar\eta\ e^{\bar\eta\cdot \xi} \langle \bar\eta|\xi\rangle
  = \int d\xi d\bar\eta\ e^{2\bar\eta\cdot \xi} =2^l~.
\end{eqnarray}

Let us end this section by deriving a few expressions involving fermionic delta functions
that are helpful in the computation of section~\ref{subs:TS}.
Using   $\tilde\psi^a_i=\sqrt{\frac{\beta}2}(\eta^a_i+\bar\lambda^a_i)$,
as defined in (\ref{tilde fermion}) (here fermions are
labelled by two indices, a tangent space index $a$ and an SO($N$) R-symmetry index $i$),
we may compute  
\begin{eqnarray}
  \int d\bar\zeta d\eta~e^{(\bar\lambda-\bar\zeta)\cdot \eta} \tilde
  \psi^a_{i}
  &=&\sqrt{\frac{\beta}{2}}\Big(\frac{\partial}{\partial\bar\lambda^{i}_a}+\bar\lambda^a_{i}
  \Big)\int d\bar\zeta d\eta~e^{(\bar\lambda-\bar\zeta)\cdot \eta}
  \nonumber\\
  &=& \sqrt{\frac{\beta}{2}}\Big(\frac{\partial}{\partial\bar\rho^{i}_a}+\bar\lambda^a_{i}
  \Big)\int d\bar\zeta~\delta(\bar\zeta-\bar\rho)\Big|_{\rho=\lambda}
\end{eqnarray}
where the above apparently baroque notation is used because the
fermionic derivatives only act upon the delta-function and not on
eventual $\bar\lambda$-dependent expression that may appear next to
it. We thus have  
\begin{eqnarray}
  \int d\bar\zeta d\eta~e^{(\bar\lambda-\bar\zeta)\cdot \eta} \tilde
  \psi^{a}\cdot\tilde
  \psi^{b} 
  &=& \beta\tilde M^{ab} \int d\bar\zeta~\delta(\bar\zeta-\bar\rho)\Big|_{\rho=\lambda}
\end{eqnarray}
with
\begin{eqnarray}
\tilde M^{ab} =
\frac12\Big(\bar\lambda^a\cdot\bar\lambda^b+\bar\lambda^a\cdot\frac{\partial}{\partial\bar
 \rho_b}-\bar\lambda^b\cdot\frac{\partial}{\partial\bar
 \rho_a} +\frac{\partial}{\partial\bar
 \rho_a}\cdot\frac{\partial}{\partial\bar
 \rho_b}\Big) \;.
\end{eqnarray}
One can then switch to the Lorentz generators
\begin{eqnarray}
M^{ab}=
\frac12\Big(\bar\lambda^a\cdot\bar\lambda^b+\bar\lambda^a\cdot\frac{\partial}{\partial\bar
 \lambda_b}-\bar\lambda^b\cdot\frac{\partial}{\partial\bar
 \lambda_a} +\frac{\partial}{\partial\bar
 \lambda_a}\cdot\frac{\partial}{\partial\bar
 \lambda_b}\Big)
\end{eqnarray}
by suitably subtracting those terms that appear when derivatives on $\bar\lambda$ act on 
eventual functions of $\bar\lambda$ that may show up before the wave function, such as 
 $M^{cd}$ in (\ref{eq:delta-TS'}).
 
%%%%%%%%%%%%%%%%%
%\section{\textbf{\emph{B$^k_l$}} coefficients} % \section{$B^k_l$ coefficients}
\section{\texorpdfstring{\boldmath$B^k_l$}{Bkl} coefficients}
\label{app:B-coeff}

The coefficients $B^k_l(x,\bar\eta,\xi)$, defined in \eqref{B
  coefficients defined}, can be computed following the strategy
described in detail in \cite{Peeters:1993vu} for $N=0,1,2$, and in \cite{Bastianelli:2010ir} for the complex $U(N|M)$ sigma model. First of all we divide the hamiltonian
\eqref{quantum H} into three pieces contributing at most two, one or
no $p$ eigenvalues
%\begin{equation}
%\begin{split}
\begin{align}
H &= H_B+H_1+H_2\;,\quad\text{where}\nonumber\\
H_B &= \frac12g^{-1/4}p_\mu g^{1/2}g^{\mu\nu}p_\nu g^{-1/4}\nonumber\\
H_1 &= -i\,g^{\mu\nu}\omega_{\mu ab}\bar \Psi^a\pu \Psi^b\,\big(g^{1/4}p_\nu g^{-1/4}\big)\nonumber\\
H_2 &= -\frac{1}{2}\,g^{-1/2}\de_\mu\big(g^{1/2}g^{\mu\nu}\omega_{\nu
  ab}\big)\bar \Psi^a\pu \Psi^b
\nonumber\\ &
-\frac{1}{2}\,\Big(g^{\mu\nu}\omega_{\mu ab}\omega_{\nu
  cd}-8\alpha\,R_{abcd}\Big)\bar \Psi^a\pu \Psi^b\bar \Psi^c\pu
\Psi^d
+V\;.
%\end{split}
%\end{equation}
\end{align}
First of all, notice that $H_B$ is precisely the usual bosonic quantum hamiltonian, carefully studied
in the literature \cite{Peeters:1993vu, Bastianelli:2006rx}.
 Let us start with $B^k_{2k}$: the only way to have $2k$
$p$ eigenvalues is from $k$ factors of $H_B$ and no commutators taken into account, giving simply
\begin{equation}
B^k_{2k}\,p^{2k}=\left(\frac{p^2}{2}\right)^k\;.
\end{equation}
For $B^k_{2k-1}$
we can have two kinds of terms. The first comes from $k$ factors of $H_B$ with one $p$ acting as a derivative; this gives the
corresponding $B^k_{2k-1}$ of the purely bosonic model,
whose computation is explained in detail in \cite{Peeters:1993vu, Bastianelli:2006rx}.
The other term comes from $k-1$
factors of $H_B$ and one $H_1$, by substituting all operators with the corresponding eigenvalues.
Putting things together we obtain
\begin{equation}
B^k_{2k-1}\,p^{2k-1} = -\frac{i
k}{2}\,\left(\frac{p^2}{2}\right)^{k-1}\!\!\!\!\!\!g^\mu\,p_\mu-i\,\binom{k}{2}\,
  \left(\frac{p^2}{2}\right)^{k-2}\,\frac12  g^{\nu\lambda\mu}\,p_\mu\,p_\nu\,p_\lambda-i\,k\,\left(\frac{p^2}{2}\right)^{k-1}\!\!\!g^{\mu\nu}\omega_\mu\,p_\nu\,.
\end{equation}
 For $B^k_{2k-2}$ four types of term
contribute: \emph{i)} $k$ factors of $H_B$, giving the coefficient of the corresponding bosonic model,
\emph{ii)} $k-1$ factors of $H_B$ and one $H_1$, with one $p$ acting as a derivative. This
contribution gives four terms: the derivative acting from one $H_B$ to $H_1$, from $H_1$ to one
$H_B$, within $H_1$ or within the $k-1$ $H_B$'s. \emph{iii)} $k-1 $ factors of $H_B$ and one $H_2$,
substituting all operators with their eigenvalues, and \emph{iv)} $k-2$ factors of $H_B$ and two
$H_1$, substituting all with eigenvalues. Remember that in \emph{iii)} and \emph{iv)} $\{\Psi,\bar\Psi\}$
anticommutators have to be taken into account in order to obtain eigenvalues on the coherent
states. Altogether it results in
\begin{align}
&B^k_{2k-2}\,p^{2k-2} =
  k\left(\frac{p^2}{2}\right)^{k-1}\Big[\frac{1}{32}\de_\mu\ln
    g\de^\mu\ln g+\frac18\de_\mu\de^\mu\ln g+\frac18g^\mu\de_\mu\ln
    g\Big]\nonumber\\&-\binom{k}{2}\left(\frac{p^2}{2}\right)^{k-2}
\Big[\frac12\de^\mu g^\nu+\frac14g^\mu g^\nu+\frac14g^\lambda
  g^{\mu\nu}_\lambda +\frac14g^{\mu\nu\lambda}_\lambda\Big]\,p_\mu\,p_\nu
\nonumber\\ &-\binom{k}{3}\left(\frac{p^2}{2}\right)^{k-3}
\Big[\frac12g^{\lambda\sigma\mu\nu}
  +\frac34g^{\mu\nu\lambda}g^\sigma
  +\frac12g^{\rho \mu\nu}g^{\lambda\sigma}_\rho+\frac14g^{\mu\nu}_\rho g^{\lambda\sigma\rho}\Big]\,p_\mu\,p_\nu\,p_\lambda\,p_\sigma\nonumber\\&-\binom{k}{4}\left(\frac{p^2}{2}\right)^{k-4}
\Big[\frac34g^{\nu\lambda\mu}g^{\rho\tau\sigma}\Big]\,p_\mu\,p_\nu\,p_\lambda\,p_\sigma\,p_\rho\,p_\tau\nonumber\\ \displaybreak
&-\binom{k}{2}\left(\frac{p^2}{2}\right)^{k-2}\Big[\de^\mu\big(g^{\nu\lambda}\omega_\lambda\big)-\frac12g^{\mu\nu\lambda}\omega_\lambda\Big]p_\mu\,p_\nu-k\Big[\frac12\,(k-1)
\left(\frac{p^2}{2}\right)^{k-2}g^\mu p_\mu\nonumber\\ 
&+\binom{k-1}{2}\left(\frac{p^2}{2}\right)^{k-3}
\frac12g^{\nu\lambda\mu}\,p_\mu\,p_\nu\,p_\lambda\Big]g^{\sigma\rho}\omega_\sigma\,p_\rho
+\frac{1}{4}k\left(\frac{p^2}{2}\right)^{k-1}g^{\mu\nu}\omega_\mu\de_\nu
\ln g\nonumber\\
&-\frac{1}{2}k\left(\frac{p^2}{2}\right)^{k-1}\Big\{g^{-1/2}\de_\mu\big(g^{1/2}g^{\mu\nu}\omega_\nu
\big)+\Big(g^{\mu\nu}\omega_{\mu
  ab}\omega_{\nu
  cd}-8\alpha\,R_{abcd}\Big)\Big[\bar\eta^a\pu\xi^d\eta^{bc}\nonumber\\
&+\bar\eta^a\pu\xi^b\bar\eta^c\pu\xi^d\Big]-2V\Big\}
\nonumber\\&-\binom{k}{2}\left(\frac{p^2}{2}\right)^{k-2}g^{\mu\nu}
\omega_{\mu ab}g^{\lambda\sigma}\omega_{\lambda
  cd}\Big[\bar\eta^a\pu\xi^d\eta^{bc}+\bar\eta^a\pu\xi^b\bar\eta^c\pu\xi^d\Big]p_\nu\,p_\sigma
\end{align}
where again we use the compact notations for tensors introduced below eq. (\ref{eq2.7}).

%%%%%%%%%%%%%%%%%%%%%%%%%%%%%%%%%%%%%%%%%%%%%%%%%%%%%%%%%%

%%%%%%%%%%%%%%%%%%%%%%%%%%%
\section{Feynman diagrams}
\label{app:FD}
We list the set of Feynman diagrams and the associated worldline
integrals that enter the computation of the transition amplitude to
order $\beta$ in section~\ref{sec:t-amp}.. We are not
reporting here those diagrams that involve fermionic self-contractions
as with the rules of section~\ref{sec:t-amp} such 
self-contrations are trivially vanishing. Hence, the a priori non-trivial diagrams
entering the contribution $\langle S_3\rangle$ are
\begin{align}
{\bf I}_1 {}=&\
\parbox{50pt}{
     \begin{fmfgraph}(30,30)
       \fmfleft{i1}
       \fmfright{o2}
       \fmf{plain}{i1,o2}
       \fmf{ddeld}{o2,o2}
   \end{fmfgraph}}\ +
\parbox{50pt}{
     \begin{fmfgraph}(30,30)
       \fmfleft{i1}
       \fmfright{o2}
       \fmf{plain}{i1,o2}
       \fmf{dashes}{o2,o2}
   \end{fmfgraph}} = \int_{-1}^0d\tau~\tau (\ddeld+\dddel)|_\tau
\\[1mm]
{\bf I}_{2} {}=&\
\parbox{50pt}{
  \begin{fmfgraph}(30,30)
    \fmfleft{i1} \fmfright{v2}
    \fmf{deld}{i1,v2}
    \fmf{ddel}{v2,v2}
\end{fmfgraph}} = \int_{-1}^0d\tau ~\ddel|_\tau
%\\
\end{align}
Those contributing to $\langle S_4\rangle$ are
%\begin{eqnarray}
\begin{align}
{\bf I}_3 {}=&\
\parbox{50pt}{
  \begin{fmfgraph}(50,30)
    \fmfleft{i}
    \fmf{plain,left}{i,o}
    \fmf{plain,right}{i,o}
    \fmfright{v}
    \fmf{deld,left}{o,v}
    \fmf{deld,right}{o,v}
\end{fmfgraph}}\ +\
\parbox{50pt}{
  \begin{fmfgraph}(50,30)
    \fmfleft{i}
    \fmf{plain,left}{i,o}
    \fmf{plain,right}{i,o}
    \fmfright{v}
    \fmf{dashes,left}{o,v}
    \fmf{dashes,right}{o,v}
\end{fmfgraph}}= \int_{-1}^0d\tau ~\del|_\tau (\ddeld+\dddel)|_\tau
\\[1mm]
{\bf I}_4 {}=&\ \parbox{50pt}{
     \begin{fmfgraph}(50,30)
       \fmfleft{i}
       \fmf{plain,left}{i,o}
       \fmf{ddel,right}{i,o}
       \fmfright{v}
       \fmf{deld,left}{o,v}
       \fmf{plain,right}{o,v}
   \end{fmfgraph}}= \int_{-1}^0d\tau ~\ddel^2|_\tau
\\[1mm]
{\bf I}_5 {}=&\
\parbox{50pt}{
  \begin{fmfgraph}(50,30)
    \fmfleft{i1,i2}
    \fmf{plain}{i1,o}
    \fmf{plain}{i2,o}
    \fmfright{v}
    \fmf{deld,left}{o,v}
    \fmf{deld,right}{o,v}
\end{fmfgraph}}\ +\
\parbox{50pt}{
  \begin{fmfgraph}(50,30)
    \fmfleft{i1,i2}
    \fmf{plain}{i1,o}
    \fmf{plain}{i2,o}
    \fmfright{v}
    \fmf{dashes,left}{o,v}
    \fmf{dashes,right}{o,v}
\end{fmfgraph}}= \int_{-1}^0d\tau ~\tau^2 (\ddeld+\dddel)|_\tau
\\[1mm]
{\bf I}_6 {}=&\
\parbox{50pt}{
  \begin{fmfgraph}(50,30)
    \fmfleft{i1,i2}
    \fmf{deld}{i1,o}
    \fmf{deld}{i2,o}
    \fmfright{v}
    \fmf{plain,left}{o,v}
    \fmf{plain,right}{o,v}
\end{fmfgraph}}= \int_{-1}^0d\tau ~\del|_\tau
\\[1mm]
{\bf I}_7 {}=&\
\parbox{50pt}{
  \begin{fmfgraph}(50,30)
    \fmfleft{i1,i2}
    \fmf{plain}{i1,o}
    \fmf{deld}{i2,o}
    \fmfright{v}
    \fmf{deld,left}{o,v}
    \fmf{plain,right}{o,v}
\end{fmfgraph}}= \int_{-1}^0d\tau ~\tau \deld|_\tau~.
\end{align}
The remaining ones contributing to $\langle S^2_3\rangle_c$ can be devided into 
purely bosonic contributions
\begin{align}
{\bf I}_9 {}=&\
   \parbox{30pt}{
     \begin{fmfgraph}(30,30)
       \fmfleft{i} \fmfright{o} \fmf{ddeld,left}{i,o}
       \fmf{ddeld,right}{i,o} \fmf{plain}{i,o}
   \end{fmfgraph}}\ +\
   \parbox{30pt}{
     \begin{fmfgraph}(30,30)
       \fmfleft{i} \fmfright{o} \fmf{dashes,left}{i,o}
       \fmf{dashes,right}{i,o} \fmf{plain}{i,o}
   \end{fmfgraph}} = \int_{-1}^0\int_{-1}^0d\tau d\sigma ~\del(\ddeld{}^2-\dddel{}^2)
\\[1mm] \displaybreak 
{\bf I}_{10} {}=&\
   \parbox{30pt}{
     \begin{fmfgraph}(30,30)
       \fmfleft{i} \fmfright{o} \fmf{ddeld,left}{i,o}
       \fmf{deld,right}{i,o} \fmf{ddel}{i,o}
   \end{fmfgraph}}=\int_{-1}^0\int_{-1}^0d\tau d\sigma ~\ddel\ddeld\deld
\\[1mm]
{\bf I}_{11} {}=&\
\left[\hskip.75cm
  \parbox{50pt}{
    \begin{fmfgraph}(30,30)
      \fmfleft{i1}
      \fmfright{o2}
      \fmf{plain}{i1,o2}
      \fmf{ddeld}{i1,i1}
  \end{fmfgraph}}\hskip-.5cm+
\hskip.75cm
  \parbox{50pt}{
    \begin{fmfgraph}(30,30)
      \fmfleft{i1}
      \fmfright{o2}
      \fmf{plain}{i1,o2}
      \fmf{dashes}{i1,i1}
  \end{fmfgraph}}\hskip-.5cm\right] \times \left[
     \parbox{50pt}{
     \begin{fmfgraph}(30,30)
       \fmfleft{i1}
       \fmfright{o2}
       \fmf{plain}{i1,o2}
       \fmf{ddeld}{o2,o2}
   \end{fmfgraph}}\ +
\parbox{50pt}{
     \begin{fmfgraph}(30,30)
       \fmfleft{i1}
       \fmfright{o2}
       \fmf{plain}{i1,o2}
       \fmf{dashes}{o2,o2}
   \end{fmfgraph}}\right]\nonumber\\
{}=&\ \int_{-1}^0\int_{-1}^0d\tau d\sigma ~(\ddeld+\dddel)|_\tau ~\del~(\ddeld+\dddel)|_\sigma
\\[1mm]
{\bf I}_{12} {}=&\  \hskip.75cm
  \parbox{50pt}{
    \begin{fmfgraph}(30,30)
      \fmfleft{i1}
      \fmfright{o2}
      \fmf{ddel}{i1,o2}
      \fmf{ddel}{i1,i1}
  \end{fmfgraph}}\hskip-.5cm
 \times \left[
     \parbox{50pt}{
     \begin{fmfgraph}(30,30)
       \fmfleft{i1}
       \fmfright{o2}
       \fmf{plain}{i1,o2}
       \fmf{ddeld}{o2,o2}
   \end{fmfgraph}}\ +
\parbox{50pt}{
     \begin{fmfgraph}(30,30)
       \fmfleft{i1}
       \fmfright{o2}
       \fmf{plain}{i1,o2}
       \fmf{dashes}{o2,o2}
   \end{fmfgraph}}\right]\nonumber\\
{}=&\ \int_{-1}^0\int_{-1}^0d\tau d\sigma ~\ddel|_\tau ~\ddel~(\ddeld+\dddel)|_\sigma
\\[1mm]
{\bf I}_{13} {}=&\  \hskip.75cm
  \parbox{50pt}{
    \begin{fmfgraph}(30,30)
      \fmfleft{i1}
      \fmfright{o2}
      \fmf{ddeld}{i1,o2}
      \fmf{ddel}{i1,i1}
      \fmf{deld}{o2,o2}
  \end{fmfgraph}}
{}=\ \int_{-1}^0\int_{-1}^0d\tau d\sigma ~\ddel|_\tau ~\ddeld~\deld|_\sigma
\\%[1mm]
{\bf I}_{14} {}=&\
   \parbox{50pt}{
     \begin{fmfgraph}(50,40)
       \fmfleft{i1} \fmfright{o1}
       \fmf{deld}{i1,v1} \fmf{ddel}{v2,o1} \fmf{plain,right,tension=.2}{v1,v2}
       \fmf{ddeld,left,tension=.2}{v1,v2}
   \end{fmfgraph}} =\int_{-1}^0\int_{-1}^0d\tau d\sigma ~\ddeld ~\del
\\[1mm]
{\bf I}_{15} {}=&\
   \parbox{50pt}{
     \begin{fmfgraph}(50,40)
       \fmfleft{i1} \fmfright{o1}
       \fmf{deld}{i1,v1} \fmf{ddel}{v2,o1} \fmf{deld,right,tension=.2}{v1,v2}
       \fmf{ddel,left,tension=.2}{v1,v2}
   \end{fmfgraph}} =\int_{-1}^0\int_{-1}^0d\tau d\sigma ~\ddel ~\deld
\\[1mm]
{\bf I}_{16} {}=&\
   \parbox{50pt}{
     \begin{fmfgraph}(50,40)
       \fmfleft{i1} \fmfright{o1}
       \fmf{plain}{i1,v1} \fmf{plain}{v2,o1} \fmf{ddeld,right,tension=.2}{v1,v2}
       \fmf{ddeld,left,tension=.2}{v1,v2}
   \end{fmfgraph}} +
\parbox{50pt}{
     \begin{fmfgraph}(50,40)
       \fmfleft{i1} \fmfright{o1}
       \fmf{plain}{i1,v1} \fmf{plain}{v2,o1} \fmf{dashes,right,tension=.2}{v1,v2}
       \fmf{dashes,left,tension=.2}{v1,v2}
   \end{fmfgraph}}=\int_{-1}^0\int_{-1}^0d\tau d\sigma ~\tau\sigma(\ddeld{}^2 -\dddel{}^2)
\\[1mm]
{\bf I}_{17} {}=&\
   \parbox{50pt}{
     \begin{fmfgraph}(50,40)
       \fmfleft{i1} \fmfright{o1}
       \fmf{plain}{i1,v1} \fmf{ddel}{v2,o1} \fmf{ddel,right,tension=.2}{v1,v2}
       \fmf{ddeld,left,tension=.2}{v1,v2}
   \end{fmfgraph}}=\int_{-1}^0\int_{-1}^0d\tau d\sigma ~\ddeld ~\ddel
\\[1mm]
{\bf I}_{18} {}=&\
   \parbox{30pt}{
     \begin{fmfgraph}(30,30)
       \fmfleft{i1,i2} \fmfright{v2}
       \fmf{deld}{i1,v1} \fmf{deld}{i2,v1} \fmf{plain}{v1,v2}
   \end{fmfgraph}}\ \left[
     \parbox{50pt}{
     \begin{fmfgraph}(30,30)
       \fmfleft{i1}
       \fmfright{e2}
       \fmf{plain}{i1,e2}
       \fmf{ddeld}{e2,e2}
   \end{fmfgraph}}\ +
\parbox{50pt}{
     \begin{fmfgraph}(30,30)
       \fmfleft{i1}
       \fmfright{e2}
       \fmf{plain}{i1,e2}
       \fmf{dashes}{e2,e2}
   \end{fmfgraph}}\right]=\int_{-1}^0\int_{-1}^0d\tau d\sigma
   ~\del~(\ddeld+\dddel)|_\sigma
\\[1mm]
{\bf I}_{19} {}=&\
   \parbox{70pt}{
     \begin{fmfgraph}(50,30)
       \fmfleft{i1,i2} \fmfright{v2}
       \fmf{deld}{i1,v1} \fmf{deld}{i2,v1} \fmf{deld}{v1,v2}
       \fmffreeze
       \fmf{ddel}{v2,v2}
   \end{fmfgraph}}=\int_{-1}^0\int_{-1}^0d\tau d\sigma
   ~\deld~\deld|_\sigma
\\[1mm]
{\bf I}_{20} {}=&\
   \parbox{30pt}{
     \begin{fmfgraph}(30,30)
       \fmfleft{i1,i2} \fmfright{v2}
       \fmf{deld}{i1,v1} \fmf{plain}{i2,v1} \fmf{ddel}{v1,v2}
   \end{fmfgraph}}\ \left[
     \parbox{50pt}{
     \begin{fmfgraph}(30,30)
       \fmfleft{i1}
       \fmfright{o2}
       \fmf{plain}{i1,o2}
       \fmf{ddeld}{o2,o2}
   \end{fmfgraph}}\ +
\parbox{50pt}{
     \begin{fmfgraph}(30,30)
       \fmfleft{i1}
       \fmfright{o2}
       \fmf{plain}{i1,o2}
       \fmf{dashes}{o2,o2}
   \end{fmfgraph}}\right]=\int_{-1}^0\int_{-1}^0d\tau d\sigma
   ~\tau ~\ddel~(\ddeld+\dddel)|_\sigma
\\[1mm]
{\bf I}_{21} {}=&\
   \parbox{70pt}{
     \begin{fmfgraph}(50,30)
       \fmfleft{i1,i2} \fmfright{v2}
       \fmf{deld}{i1,v1} \fmf{plain}{i2,v1} \fmf{ddeld}{v1,v2}
       \fmffreeze
       \fmf{ddel}{v2,v2}
   \end{fmfgraph}}=\int_{-1}^0\int_{-1}^0d\tau d\sigma
   ~\tau~\ddeld~\deld|_\sigma
\\[1mm]
{\bf I}_{22} {}=&\
   \parbox{50pt}{
     \begin{fmfgraph}(50,30)
       \fmfleft{i1,i2} \fmfright{o1,o2}
       \fmf{deld}{i1,v1} \fmf{deld}{i2,v1} \fmf{plain}{v1,v2}
       \fmf{ddel}{v2,o1} \fmf{ddel}{v2,o2}
   \end{fmfgraph}}=\int_{-1}^0\int_{-1}^0d\tau d\sigma
   ~\del
\\[1mm]
{\bf I}_{23} {}=&\
   \parbox{50pt}{
     \begin{fmfgraph}(50,30)
       \fmfleft{i1,i2} \fmfright{o1,o2}
       \fmf{deld}{i1,v1} \fmf{deld}{i2,v1} \fmf{deld}{v1,v2}
       \fmf{ddel}{v2,o1} \fmf{plain}{v2,o2}
   \end{fmfgraph}}=\int_{-1}^0\int_{-1}^0d\tau d\sigma
   ~\sigma~\deld
\\[1mm]
{\bf I}_{24} {}=&\
   \parbox{50pt}{
     \begin{fmfgraph}(50,30)
       \fmfleft{i1,i2} \fmfright{o1,o2}
       \fmf{deld}{i1,v1} \fmf{plain}{i2,v1} \fmf{ddeld}{v1,v2}
       \fmf{ddel}{v2,o1} \fmf{plain}{v2,o2}
   \end{fmfgraph}}=\int_{-1}^0\int_{-1}^0d\tau d\sigma
   ~\tau\sigma~\ddeld
\end{align}
\vfill\eject
and those with mixed bosonic-fermionic contributions
\begin{align}
{\bf I}_{25} {}=&\
   \parbox{30pt}{
     \begin{fmfgraph}(30,30)
       \fmfleft{i} \fmfright{o}
       \fmf{ddeld}{i,o} \fmf{dots,left}{i,o} \fmf{dots,right}{i,o}
   \end{fmfgraph}} = \int_{-1}^0\int_{-1}^0d\tau d\sigma
   ~\ddeld~\del_F^2
\\[1mm]
{\bf I}_{26} {}=&\
   \parbox{50pt}{
     \begin{fmfgraph}(50,40)
       \fmfleft{i1} \fmfright{o1}
       \fmf{deld}{i1,v1} \fmf{ddel}{v2,o1} \fmf{dots,right,tension=.2}{v1,v2}
       \fmf{dots,left,tension=.2}{v1,v2}
   \end{fmfgraph}}= \int_{-1}^0\int_{-1}^0d\tau d\sigma
   ~\del_F^2
\\[1mm]
{\bf I}_{27} {}=&\
   \parbox{50pt}{
     \begin{fmfgraph}(50,40)
       \fmfleft{i1} \fmfright{o1}
       \fmf{dots}{i1,v1} \fmf{dots}{v2,o1} \fmf{dots,right,tension=.2}{v1,v2}
       \fmf{ddeld,left,tension=.2}{v1,v2}
   \end{fmfgraph}}= \int_{-1}^0\int_{-1}^0d\tau d\sigma
   ~\ddeld~\del_F
\\%[1mm]
{\bf I}_{28} {}=&\
   \parbox{50pt}{
     \begin{fmfgraph}(50,30)
       \fmfleft{i1,i2} \fmfright{o1,o2}
       \fmf{deld}{i1,v1} \fmf{dots}{i2,v1,v2,o2}
       \fmf{ddel}{v2,o1}
   \end{fmfgraph}}= \int_{-1}^0\int_{-1}^0d\tau d\sigma
   ~\del_F
\\[1mm]
{\bf I}_{29} {}=&\
   \parbox{50pt}{
     \begin{fmfgraph}(50,30)
       \fmfleft{i1,i2} \fmfright{o1,o2}
       \fmf{dots}{i1,v1,i2} \fmf{ddeld}{v1,v2}
       \fmf{dots}{o1,v2,o2}
   \end{fmfgraph}}= \int_{-1}^0\int_{-1}^0d\tau d\sigma
   ~\ddeld
\\[1mm]
{\bf I}_{30} {}=&\
   \parbox{30pt}{
     \begin{fmfgraph}(30,30)
       \fmfleft{i1,i2} \fmfright{v2} %\fmf{ddeld,left}{i,o}
       \fmf{dots}{i1,v1,i2} \fmf{ddel}{v1,v2}
   \end{fmfgraph}}\ \left[
     \parbox{50pt}{
     \begin{fmfgraph}(30,30)
       \fmfleft{i1}
       \fmfright{o2} %\fmf{ddeld,left}{i,o}
       \fmf{plain}{i1,o2}
       \fmf{ddeld}{o2,o2}
   \end{fmfgraph}}\ +
\parbox{50pt}{
     \begin{fmfgraph}(30,30)
       \fmfleft{i1}
       \fmfright{o2} %\fmf{ddeld,left}{i,o}
       \fmf{plain}{i1,o2}
       \fmf{dashes}{o2,o2}
   \end{fmfgraph}}\right]= \int_{-1}^0\int_{-1}^0d\tau d\sigma
   ~\ddel~(\ddeld+\dddel)|_\sigma
\\[1mm]%[1mm]
{\bf I}_{31} {}=&\
   \parbox{70pt}{
     \begin{fmfgraph}(50,30)
       \fmfleft{i1,i2} \fmfright{v2} %\fmf{ddeld,left}{i,o}
       \fmf{dots}{i1,v1,i2}\fmf{ddeld}{v1,v2}
       \fmffreeze
       \fmf{ddel}{v2,v2}
   \end{fmfgraph}}= \int_{-1}^0\int_{-1}^0d\tau d\sigma
   ~\ddeld~\deld|_\sigma
\\[1mm]%[1mm]
{\bf I}_{32} {}=&\
   \parbox{50pt}{
     \begin{fmfgraph}(50,30)
       \fmfleft{i1,i2} \fmfright{o1,o2} %\fmf{ddeld,left}{i,o}
       \fmf{dots}{i1,v1,i2} \fmf{ddel}{v1,v2}
       \fmf{ddel}{v2,o1} \fmf{ddel}{v2,o2}
   \end{fmfgraph}}= \int_{-1}^0\int_{-1}^0d\tau d\sigma
   ~\ddel
\\[1mm]%[1mm]
{\bf I}_{33} {}=&\
   \parbox{50pt}{
     \begin{fmfgraph}(50,30)
       \fmfleft{i1,i2} \fmfright{o1,o2} %\fmf{ddeld,left}{i,o}
       \fmf{dots}{i1,v1,i2} \fmf{ddeld}{v1,v2}
       \fmf{plain}{v2,o1} \fmf{ddel}{v2,o2}
   \end{fmfgraph}}= \int_{-1}^0\int_{-1}^0d\tau d\sigma
   ~\sigma~\ddeld
%\end{eqnarray}
\end{align}
}
\end{fmffile}

%%%%%%%%%%%%%%%%%%%%%%%%%%%%%%%%%%%%%%%%%%%%%%%%%%%%%%%%
%%%%%%%%%%%%%%%%%%%%%%%%%%%%%%%%%%%%%%%%%%%%%%%%%%%%%%%%%%

\end{document}